\documentclass[aps,prb,twocolumn,superscriptaddress,showpacs]{revtex4-1}

\usepackage{ulem}
\usepackage[utf8]{inputenc}
\usepackage{graphicx}
\usepackage{dcolumn}
\usepackage{bm}
\usepackage{hyperref}
\usepackage{amsmath}
\usepackage{mathtools}
\usepackage[dvipsnames]{xcolor}

\begin{document}

	\preprint{APS/123-QED}

	\title{Disordered hyperuniform vortex matter with rhombic distortions
		in FeSe at low fields}

	\author{Jazm\'{i}n Arag\'{o}n S\'{a}nchez}%
	\affiliation{Centro At\'{o}mico Bariloche and Instituto Balseiro,
		CNEA, CONICET and Universidad Nacional de Cuyo, Avenida Bustillo 9500, 8400 San Carlos de
		Bariloche, Argentina}
	\affiliation{Leibniz Institute for Solid State and Materials Research, Helmholtzstra$\beta$e 20, 01069 Dresden, Germany}
	
	\author{Ra\'ul Cort\'es Maldonado}
	\affiliation{Centro At\'{o}mico Bariloche and Instituto Balseiro,
		CNEA, CONICET and Universidad Nacional de Cuyo, Avenida Bustillo 9500, 8400 San Carlos de
		Bariloche, Argentina}

	\author{M. Lourdes Amig\'o}
	\affiliation{Centro At\'{o}mico Bariloche and Instituto Balseiro,
		CNEA, CONICET and Universidad Nacional de Cuyo, Avenida Bustillo 9500, 8400 San Carlos de
		Bariloche, Argentina}
	
	\author{Gladys Nieva}
	\affiliation{Centro At\'{o}mico Bariloche and Instituto Balseiro,
		CNEA, CONICET and Universidad Nacional de Cuyo, Avenida Bustillo 9500, 8400 San Carlos de
		Bariloche, Argentina}
	
	\author{Alejandro Kolton}
	\affiliation{Centro At\'{o}mico Bariloche and Instituto Balseiro,
		CNEA, CONICET and Universidad Nacional de Cuyo, Avenida Bustillo 9500, 8400 San Carlos de
		Bariloche, Argentina}

	\author{Yanina Fasano}
	\affiliation{Centro At\'{o}mico Bariloche and Instituto Balseiro,
		CNEA, CONICET and Universidad Nacional de Cuyo, Avenida Bustillo 9500, 8400 San Carlos de
		Bariloche, Argentina}
	\affiliation{Leibniz Institute for Solid State and Materials Research, Helmholtzstra$\beta$e 20, 01069 Dresden, Germany}
	
	\date{\today}

	\begin{abstract}

		In the current quest to synthesize hyperuniform materials with  interesting applications, addressing the coupling of the objects composing the system  to the physical properties of the host medium is crucial. With this aim we study a model system: vortices in the FeSe superconductor subject to a considerable magneto-elastic coupling with the host sample. We reveal that the low-field FeSe vortex structure is of the weakest hyperuniform type possibly due to the relevance of the anisotropic and long-ranged  interaction term introduced by the magneto-elastic coupling.  This work indicates that it is possible to tailor the hyperuniformity class  of  material systems by tuning the coupling of interacting objects with elastic properties of the host medium.

	\end{abstract}
	
	\maketitle
	
\section{Introduction}	
	
	Hyperuniform structures
	are attracting much attention in applied and basic research due to
	the search of novel disordered systems with promising
	properties for applications~\cite{Torquato2003,Zachary2011,Man2013,Dreyfus2015, Chen2018,Rumi2019,Llorens2020,Zheng2020,Salvalaglio2020,Chieco2021,Chen2021,Chen2021b}
	such as complete photonic bandgap materials~\cite{Man2013} and highly-efficient vortex pinning structures in superconductors.~\cite{Lethien2017}
	This topic is at the crossroads of several research fields on structural properties of systems in condensed matter physics, material science, biology, mathematics, cosmology, out of equilibrium phenomena and technological devices.
	This ubiquitous
	state of matter,  characterized by vanishing density
	fluctuations at infinite wavelengths or small wavenumbers,~\cite{Torquato2018} exhibits novel  electrical, optical and structural properties.~\cite{Man2013,Chen2018,Torquato2018,Sheremet2020} Hyperuniform material systems composed of interacting objects present a homogeneous density of constituents at large scales like in a crystal, but they can be also isotropic and disordered. Mastering the synthesis of hyperuniform materials requires to assess the effect of quenched disorder in the host medium where objects are nucleated: Disorder can induce imperfections  affecting or even destroying hyperuniformity.~\cite{Kim2018,Puig2022} Furthermore, the coupling of the objects with electronic, magnetic or elastic properties of the host
	medium  can drive  structural distortions
	affecting the amount of suppression of density fluctuations known as hyperuniformity-class of the system.

	Accessing to the long-wavelength density fluctuations in extended fields-of-view is mandatory to understand these issues since it allows studying the small-wavenumber limit of
	the structure factor of the system $S(\mathbf{q})$.
	Hyperuniform systems present a vanishing $S(\mathbf{q})
	\sim q^{\alpha}$ when $\mathbf{q} \to \mathbf{0}$,
	with  $\alpha$ characterizing the hyperuniformity-class of the system:
	$\alpha > 1$ for class-I, $\alpha=1$ for class-II and $0 < \alpha <1$ for the weakest class-III.~\cite{Torquato2018}
	We follow this approach using  vortex matter in type-II superconducting samples as a toy model system since these materials are host media with different types of quenched disorder and couplings with the vortex structures. The nature and magnitude of disorder is provided by the sample defects acting as pinning centers for vortices. Despite the apparent disorder of the glassy vortex phases,~\cite{Fasano2005,Petrovic2009,AragonSanchez2019} class-II hyperuniform structures are nucleated in host media with weak
	point defects.~\cite{Rumi2019} In contrast, the vortex structure in samples with strong point defects is class-III  hyperuniform~\cite{Llorens2020} and lacks hyperuniformity in samples with planar disorder.~\cite{Puig2022}  Some superconductors have peculiar electronic and elastic properties that induce symmetry changes in the vortex structure. ~\cite{Eskildsen2011,Olszewski2018,Putilov2019,Zhang2019b,Kogan2013,Lu2018} How this affects the hyperuniform properties of the system  has not yet been elucidated.

	We study this issue in FeSe, a host medium presenting an interplay of nematic electronic order,~\cite{Fang2008,Chowdhury2011,Tanatar2016,Baek2015} non-local crystal-vortex structure interaction~\cite{Kogan1997} and magneto-elastic~\cite{Margadonna2008,Medvedev2009,Lin2017} effects.
	The  nematic order induces an elliptic spectroscopic vortex halo in FeSe at high fields~\cite{Song2011,Chowdhury2011,Watashige2015,Hanaguri2019,Putilov2019} and can drive a structural transition to an oblique lattice.~\cite{Lu2018} A hexagonal to rhombic transition is indeed observed at high vortex densities and was explained considering the multiband superconductivity of FeSe.~\cite{Putilov2019} Nevertheless, this transition can also be driven by non-local  effects~\cite{Putilov2019}: The coupling of the vortex system with the shape of the Fermi surface yields a non-local anisotropic vortex interaction.~\cite{Kogan1997} A hexagonal to oblique vortex structure transition is predicted on increasing field for nematic and non-local electronic couplings.~\cite{Lu2018,Kogan1997} In contrast,  rhombic  distortions  are expected at all fields if the superconductor has a strong magneto-elastic coupling with vortices.~\cite{Kogan2013} This effect is relevant for materials with a strong critical temperature $T_{\rm c}$ \textit{vs.} pressure $P$ dependence as FeSe.~\cite{Lin2017,Margadonna2008,Medvedev2009} This coupling adds to the short-range London vortex interaction a long-range anisotropic term given by the crystal symmetry of the sample.~\cite{Lin2017} FeSe is thus a rich playground to elucidate how the coupling with electronic and elastic properties of the media affects hyperuniformity.

In this work we present experimental  as well as simulation results of the vortex structure nucleated in FeSe. By imaging vortices in extended fields of view, we reveal that at low fields the vortex structure in FeSe is disordered but presents hexagonal symmetry with rhombic distortions. Performing  Langevin dynamics simulations of the vortex structure in FeSe we provide evidence that these distortions are quite likely produced by the magneto-elastic effect that introduces a weak, yet anisotropic and long-range, extra term in the vortex-vortex interaction. In addition, the polycrystalline vortex structure actually displays a  disordered hyperuniform order and presents a quasi-long range orientational order locked to the  symmetry axis of the magneto-elastic interaction (determined by the  crystal structure of the material). In a more general scenario, we show that if a system of elastic interacting objects is coupled to a host medium  that induces an anisotropic long-range interaction between the objects, instead of being ordered hyperuniform~\cite{Rumi2019} the system becomes disordered (class-III) hyperuniform.

\section{Experimental}	

\subsection{Magnetic decoration vortex imaging}
	
	We access to the small-$\mathbf{q}$ limit of $S(\mathbf{q})$ by imaging thousands of individual vortices via magnetic decoration.~\cite{Fasano1999} We take snapshots of the magnetic halo of vortices by evaporating ferromagnetic particles that are attracted towards the local field gradient entailed by a vortex when impinging in the sample surface. This technique discerns individual vortex positions at low fields $B\,\propto\,1/a_{0}$, with $a_{0}$ the average vortex spacing. We succedded on imaging vortices in FeSe only below $\sim 5$\,G ($a_{0}$=$2.2$\,$\mu$m) after field-cooling down to 2.3\,K. During this process the structure gets frozen at lengthscales of
	$a_{0}$ at a temperature close to the irreversibility one where  pinning sets in.~\cite{CejasBolecek2016} For the studied samples this temperature is $\sim 8.1$\,K at 5\,G, as obtained from the collapse of the field- and zero-field-cooling magnetization. On further cooling, vortices profit from disorder making excursions on lengthscales of the coherence length, $10^{-2}$  times smaller than the spatial resolution of the decoration technique $\sim\lambda(8.1$\,K$)$=$0.5$\,$\mu$m$\sim 0.2 a_{0}$. Thus,  the snapshots taken at 2.3\,K reveals the structural properties of the system at the freezing temperature. Details on the sample growth and characterization can be found in Appendix A.

\subsection{Langevin dynamics simulations}	

\begin{figure}[bbb]
    \centering
    \includegraphics[width=\columnwidth]{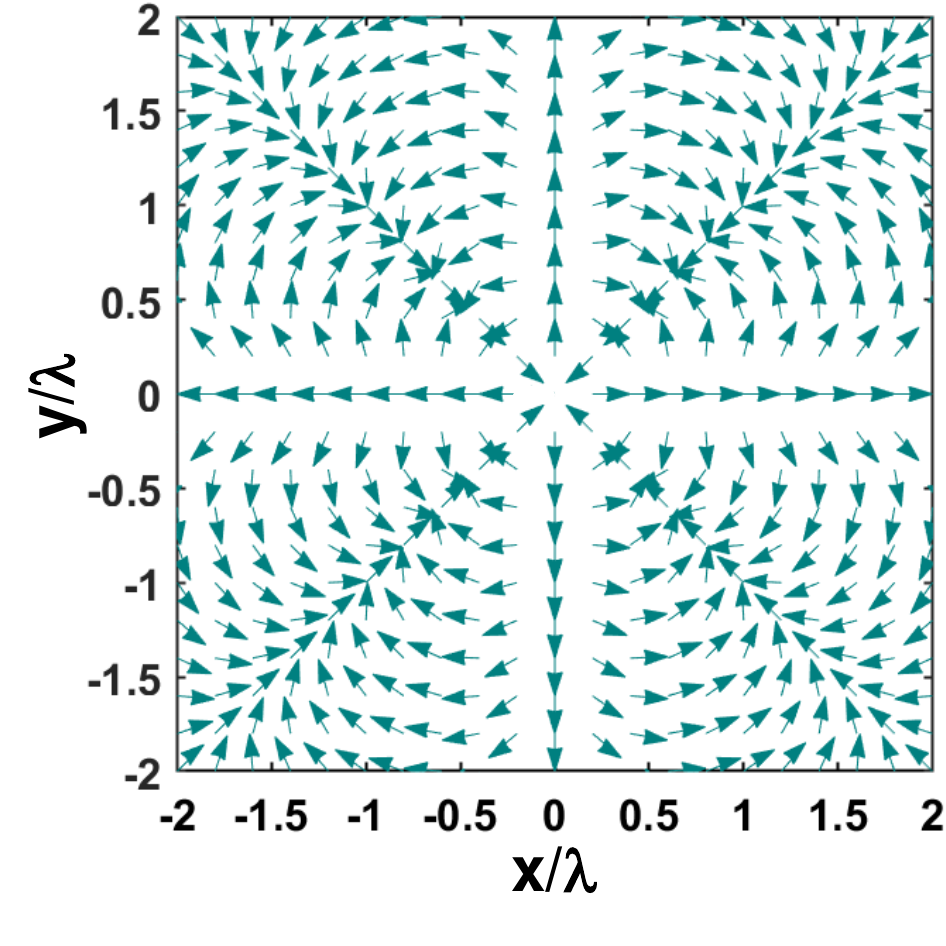}
    \caption{Normalized magneto-elastic interaction force-field exerted by a vortex located at the origin.}
    \label{fig:11}
\end{figure}
 To quantify the effect of the extra term in the vortex-vortex interaction introduced by the magneto-elastic coupling of vortices nucleated in FeSe,  we perform finite-temperature Langevin dynamics two-dimensional simulations considering this term in the approximation of Kogan \textit{et al}.~\cite{Kogan2013,Lin2017}
Although in three-dimensional vortex structures two-dimensional simulations are exact only in the case of rigid vortices, if vortex fluctuations along the $\mathbf{c}$ axis direction are weak enough compared to $\lambda$, two-dimensional simulations provide an effective model to understand the structural properties of the top layer of three-dimensional vortex matter.~\cite{Rumi2019} Two dimensional simulations have the computational advantage that a relatively large number of vortices, comparable to the typical number of vortices revealed by magnetic decorations in extended fields of view, can be achieved. We thus consider a system of $N$ overdamped straight vortices at positions ${\bf r}_i(t)$ in two dimensions with equation of motion
\begin{eqnarray}
\alpha_{\rm BS} \frac{d{\bf r}_i}{dt}= \sum_{j\neq i} {\bf F}({\bf r}_i-{\bf r}_j)+\zeta_i(t).
\label{eq:langevin}
\end{eqnarray}
Here $\alpha_{\rm BS}$ is the Bardeen-Stephen friction and $\zeta_i(t)={\hat x} \zeta_i^x(t) + {\hat y} \zeta_i^y(t)$ an isotropic Langevin noise with $\langle \zeta^\gamma_i(t) \rangle=0$ and $\langle \zeta^\gamma_i(t)\zeta^{\gamma'}_j(t') \rangle=2 \alpha_{\rm BS} k_B {\cal T}_{\rm eff} \delta_{ij}\delta_{\gamma \gamma'}\delta(t-t')$. The latter is introduced to produce steady-state configurations with different degrees of structural disorder as to model the effects of temperature and point quenched disorder in the experiments. Quenched disorder is not explicitly considered in the simulations since for low vortex densities its effect is just to slow down the dynamics during cooling. This issue is discussed in detail in Appendix G.
The noise amplitude is controlled by the effective temperature per unit length, ${\cal T}_{\rm eff}$.
We consider a pair interaction force
${\bf F}({\bf r})= -\partial_{\bf r}U$ that derives from the Kogan pair interaction potential~\cite{Kogan2013,Lin2017} for a tetragonal superconductor with magneto-elastic coupling
\begin{eqnarray}
\frac{U(r,\phi)}{\epsilon_{0}} \approx  K_{0}\left(\frac{r}{\lambda}\right) + \eta \frac{\cos{(4\phi)}}{r^2}.
\end{eqnarray}

\noindent $\epsilon_0$ is the vortex energy per unit length and $\eta$ quantifies the relative strength of the anisotropic long-ranged magneto-elastic interaction.  Even though in the superconducting phase FeSe has an orthorhombic structure, the departure from tetragonality is tiny since $a/b \sim 1.003$, and the Kogan model is thus an acceptable approach to introduce anisotropy and simulate the main structural properties of the system.
In this approximation the interaction force between a pair of vortices separated a displacement vector $\mathbf{r} = (x,y)$ reads

\begin{eqnarray}
\frac{{\bf F}(r,\phi)}{\epsilon_0}=\frac{{\bf r}}{r} K_1\left(\frac{r}{\lambda}\right)+
\eta{\bf f}(r,\phi)
\label{eq:koganmodelforces}
\end{eqnarray}
whith $\phi = \arctan(y/x)$ for the two-dimensional coordinate system aligned with the $a$ crystalline axis of the sample. The in-plane components of the extra force  introduced by the magneto-elastic effect are

\begin{eqnarray}
{f_x(r,\phi)}= \frac{2}{r^4} \left[
x \cos(4\phi)-2 y \sin(4\phi)
\right] \\
{f_y(r,\phi)}= \frac{2}{r^4} \left[
y \cos(4\phi)+2 x \sin(4\phi)
\right]
\label{eq:koganmodelforcesmagnetoelastic}
\end{eqnarray}

\noindent The normalized force-field, ${\bf f}({\bf r})=(f_x,f_y)/\sqrt{f_x^2+f_y^2}$ arising  from
the magneto-elastic interaction term for a vortex located at the origin is shown in Figure\,\ref{fig:11}.
Vectors are normalized for clarity, but the strength of the interaction forces decays as $r^{-3}$. Attractive and repulsive angular sectors are appreciated. In particular, the $\phi=2n\pi/4$ directions are repulsive while the $\phi=(2n+1)\pi/4$  are attractive for $n$ integer.

We adopt periodic boundary conditions and simulate $N=8192$ vortices with average spacing $a_{0} = 5 \lambda(T_{\rm irr})$, both magnitudes close to the experimental situation. In order to sample different configurations we let the system relax from a high effective temperature to a target one at which the total interaction energy reaches a plateau. We analyze the structure factors and Delaunay triangulations of the configurations for different values of  ${\cal T}_{\rm eff}$ and $\eta$.

	\begin{figure}[bbb]
		\centering
		\includegraphics[width=\columnwidth]{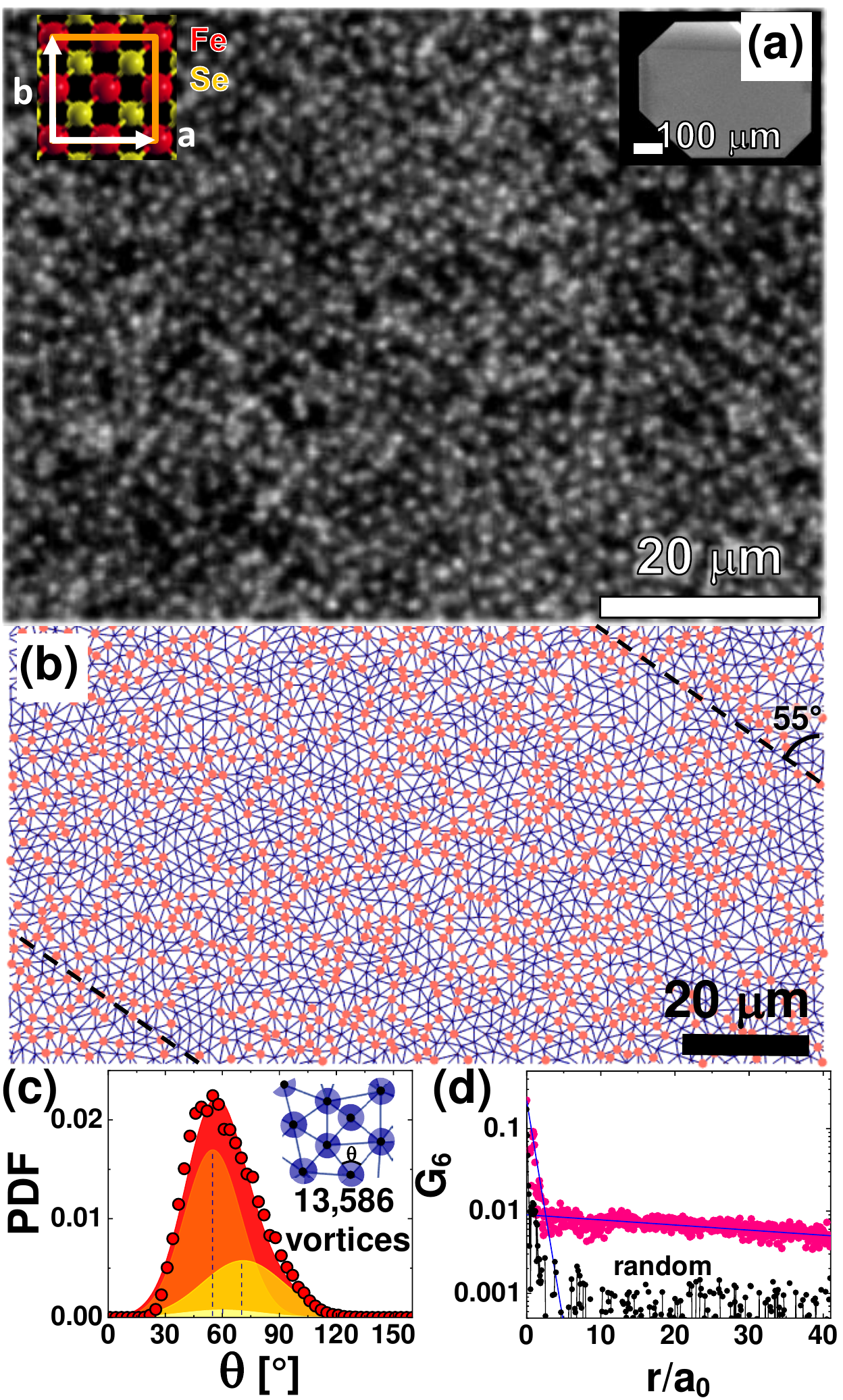}
		\caption{Field-cooled vortex structure nucleated in FeSe at 5\,G. (a) Magnetic decoration image of vortices (white dots) taken at 2.3\,K in a sample with horizontal edges along the Fe-Fe bond direction $\mathbf{a}$ of the orthorhombic (low temperature) crystal structure. We can not ascertain whether $\mathbf{a}$ is the shortest or largest vector since XR measurements were performed in the tetragonal phase. (b) Delaunay Triangulation: First-neighbours joined with blue lines and  non-sixfold coordinated vortices ($50$\,\%) highlighted in red. (c) Probability density function  of the internal angles of the triangles (circles). Red area: Sum of Gaussian distributions centered at $55$\,(orange), $70$\,(yellow) and $60^{\circ}$\,(light yellow) with respective weights of 60, 30 and 10\,\%. (d) Orientational correlation function of the  structure  (pink) compared with that of a random distribution. }
		\label{fig:1}
	\end{figure}
	
\section{Results}

\subsection{Polycrystalline vortex structure with rhombic distortions nucleated in FeSe}
 \begin{figure}[bbb]
		\centering
		\includegraphics[width=\columnwidth]{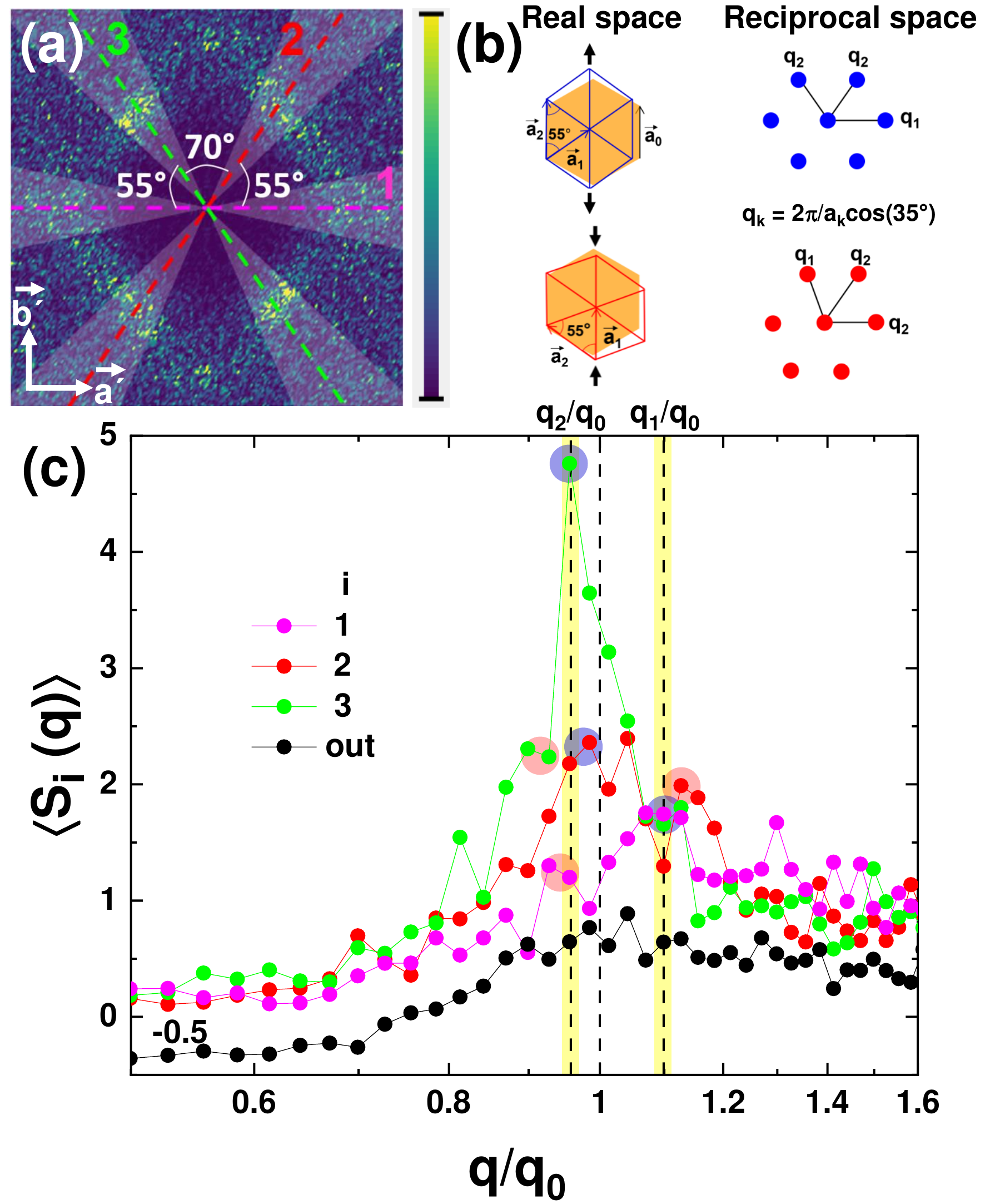}
		\caption{(a) Structure factor $S(\mathbf{q})$ of the FeSe vortex structure nucleated at 5\,G.  Angular sections 1, 2 and 3 (highlighted) considered for computing the partial angular averages  $\langle S_{\rm i} ({\bf q}) \rangle $. (b)  Left: Oblique (blue and red) and hexagonal (orange) unit cells observed in real space in different crystallites. Right: Corresponding structure factors. Arrows indicate the strain applied to the hexagons as to obtain the observed oblique cells.  Internal angles, lattice vectors and wave vectors are indicated. (c) $\langle S_{\rm i} ({\bf q}) \rangle $ data in the vicinity of the peaks. The black curve corresponds to the average of $S(\mathbf{q})$ in the regions outside sections 1, 2 and 3. Blue and red circles highlight the two local maxima coming from the blue and red oblique cells.}
		\label{fig:2}
	\end{figure}

Figure\,\ref{fig:1}\,(a) shows a zoom-in image
 of the 13586 vortices (white dots) nucleated at 5\,G and imaged at the surface of one of the studied samples.   The bottom edge of this crystal is aligned along one of the Fe-Fe bond directions and no twin boundaries were revealed by decoration in the whole sample. Results are similar in a smaller sample of the same batch with  no twins detected, see Appendix B. We studied 30 samples of the same batch and only two were untwinned (see Appendix C for data in twinned samples). Samples being twinned or untwinned may have origin in variations on the
	uniaxial stress level while cooling throughout the tetragonal to orthorhombic transition. Here we focus on samples with no detected twin boundaries and thus disorder in the host media is point-like. The vortex structure is rather disordered. Delaunay triangulation analysis reveals 50\,\% of non-sixfold coordinated vortices and crystallites with $\sim$10 sixfold-coordinated vortices (blue regions in Fig.\,\ref{fig:1}\,(b)). Density fluctuations are moderate at short lengthscales: The variance in the first neighbor's distance is $0.2 a_{0}$. The probability density function of the angles of the Delaunay triangles is  fitted by  two dominant Gaussian contributions centered at $55$ and $70\,^{\circ}$, plus a small contribution at $60\,^{\circ}$, see Fig.\,\ref{fig:1}\,(c). Thus, the low-field FeSe vortex structure has rhombic distortions.
	The orientational correlation function $G_{6}(r)$~\cite{Fasano2005} has a two-step exponential decay. A fast decay is followed by a slower one with characteristic distances $\sim\, a_{0}$ and $\sim 70\, a_{0}$, see Fig.\,\ref{fig:1}\,(d).   This weak orientational order persisting in the policrystalline structure is consistent with the observation of vortex rows  at 55\,$^{\circ}$ from the $\mathbf{b}$-axis.

	These distortions and orientational order are also evident in the $S(\mathbf{q})$ data of Fig.\,\ref{fig:2} presenting the diffraction pattern of a rhombic or likewise uniaxially-distorted hexagonal structure. We computed $S(\mathbf{q})\equiv
	S(q_{\rm x},q_{\rm y})=|\hat{\rho}(q_{\rm x},q_{\rm y},z=0)| ^{2}$,
	with $\hat{\rho}$ the Fourier transform  of the local vortex density modulation.~\cite{AragonSanchez2019}
	The broad maxima result from the positional disorder of the structure, but the six spots of a distorted hexagon are evident. The angles between the spots 1, 2 and 3 in  Fig.\,\ref{fig:2}  (a) indicate that most  crystallites present a  distorted hexagonal symmetry with a triangular unit cell with two  angles of 55 and one of 70\,$^{\circ}$. Each diffraction peak is broad since collects the signal from domains either expanded or compressed along the $\mathbf{b}$ Fe-Fe bond direction, see blue or red-type domains in Fig.\,\ref{fig:2}  (b). In this figure the orange isotropic hexagons with lattice spacing $a_{0}$ have the same area than the two distorted hexagons (expanded and compressed) with internal angles as revealed by the $S(\mathbf{q})$ data. The unit cell vectors of the distorted hexagons have modulii $a_{1}$=$0.956 a_{0}$ and $a_{2}$=$1.1 a_{0}$, with  $\mathbf{a_{1}}$ at 55\,$^{\circ}$ (0\,$^{\circ}$) from $\mathbf{b}$  in the blue (red) hexagon. Figure\,\ref{fig:2} (c) shows that the mixture of blue and red hexagons yields local maxima in the angularly-disaggregated $\langle S_{\rm i}({\bf q}) \rangle$ data. These maxima detected at $q_{1}$   and $q_{2}$ are highlighted with blue/red circles associated with the spots from expanded/compressed hexagons. The average of $S(\mathbf{q})$ outside the 1 to 3 areas has a broad peak around $q_{0}$, the Bragg wavevector of a hexagonal lattice with spacing $a_{0}$, due to the small contribution of the domains of isotropic hexagons.

\subsection{Class-III hyperuniform vortex matter in FeSe}

 \begin{figure}[ttt]
		\centering
		\includegraphics[width=0.75\columnwidth]{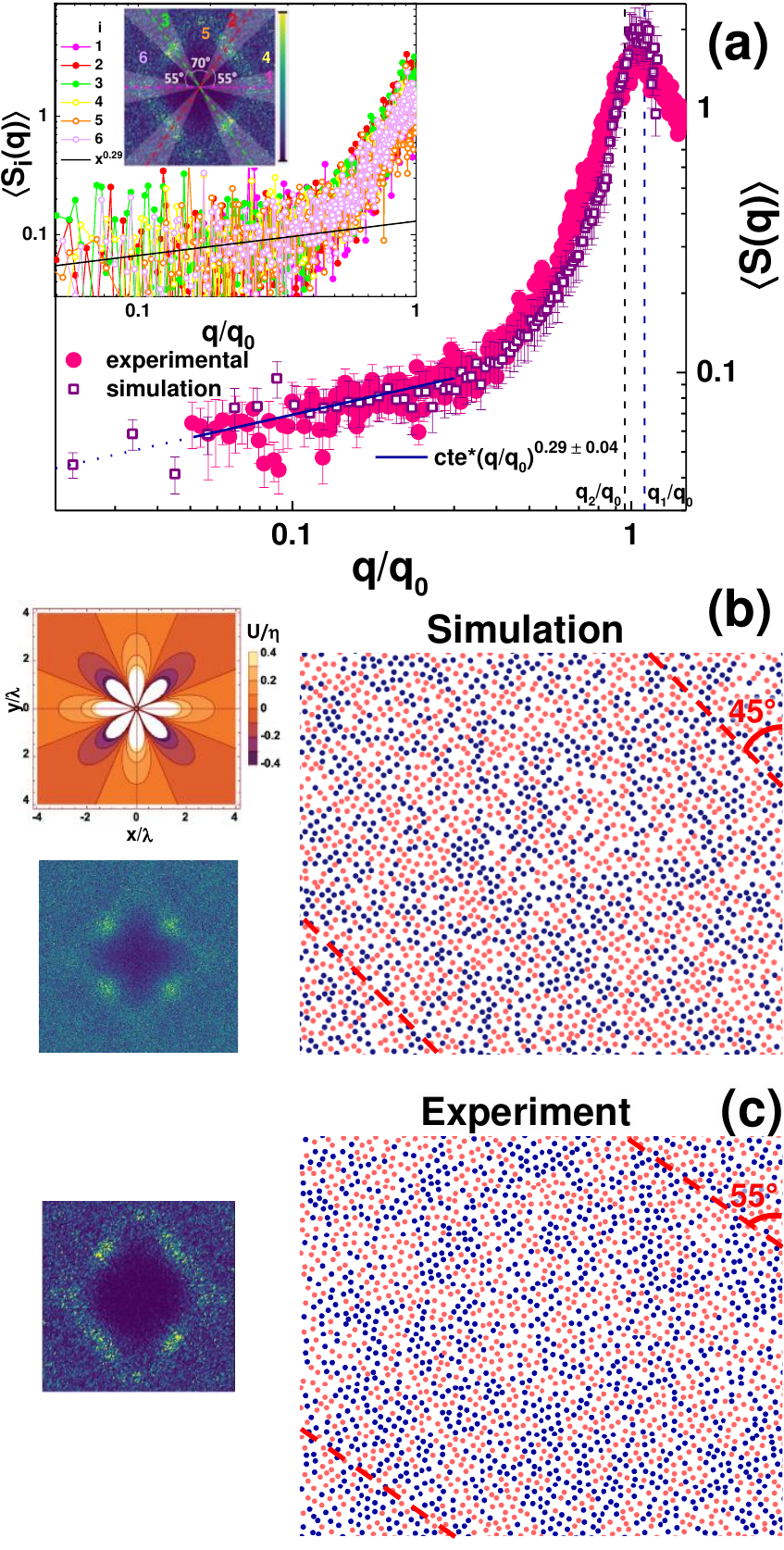}
		\caption{(a) Angularly-averaged structure factor $\langle S({\bf q}) \rangle $ of the FeSe vortex structure nucleated at 5\,G obtained experimentally and  from simulations considering a magneto-elastic coupling. In the low-$q$ limit the data follow an algebraic growing with an exponent $\alpha=0.29 \pm 0.04$ (blue line). Insert: Partial angular averages of the structure factor in regions $1$ to $6$ indicated in the $S(\mathbf{q})$.  (b) Top-left: Spatial dependence of the pair interaction potential of the long-ranged anisotropic magneto-elastic interaction in a tetragonal superconductor. Bottom-left: Structure factor of the simulated structure. Right: Typical snapshot obtained from Langevin dynamic simulations using the London plus magneto-elastic interactions between vortices. (c) Left: Structure  factor of the experimental structure. Right: Zoom in of the experimental snapshot in a region covering the same area than the results shown in (b). Non-sixfold (sixfold) coordinated vortices are presented in red (blue). The overall orientations of the simulated (experimental) vortex structure at 45\,$^{\circ}$ (55\,$^{\circ}$) from the $\mathbf{b}$-axis (vertical direction) are indicated.}
		\label{fig:3}
	\end{figure}
	
	We study the nature of long wavelength vortex density fluctuations by computing the angularly-averaged structure factor $\langle S({\bf q}) \rangle$, see Fig.\,\ref{fig:3}(a). This magnitude algebraically decays to zero when $q \rightarrow 0$ with a dependence $\propto$$(q/q_{0})^{\alpha}$ with $\alpha$=$0.29 \pm 0.04$ (see blue line). This exponent is quantitatively supported by the growth with distance of the vortex number variance, see Appendix D. Thus, despite the low field FeSe vortex matter being disordered, the fit to the data of Fig.\,\ref{fig:3}(a) is a clear indication that  it posses the hidden order of \textit{disordered} hyperuniformity.

\begin{figure}[ttt]
    \centering
    \includegraphics[width=0.75\columnwidth]{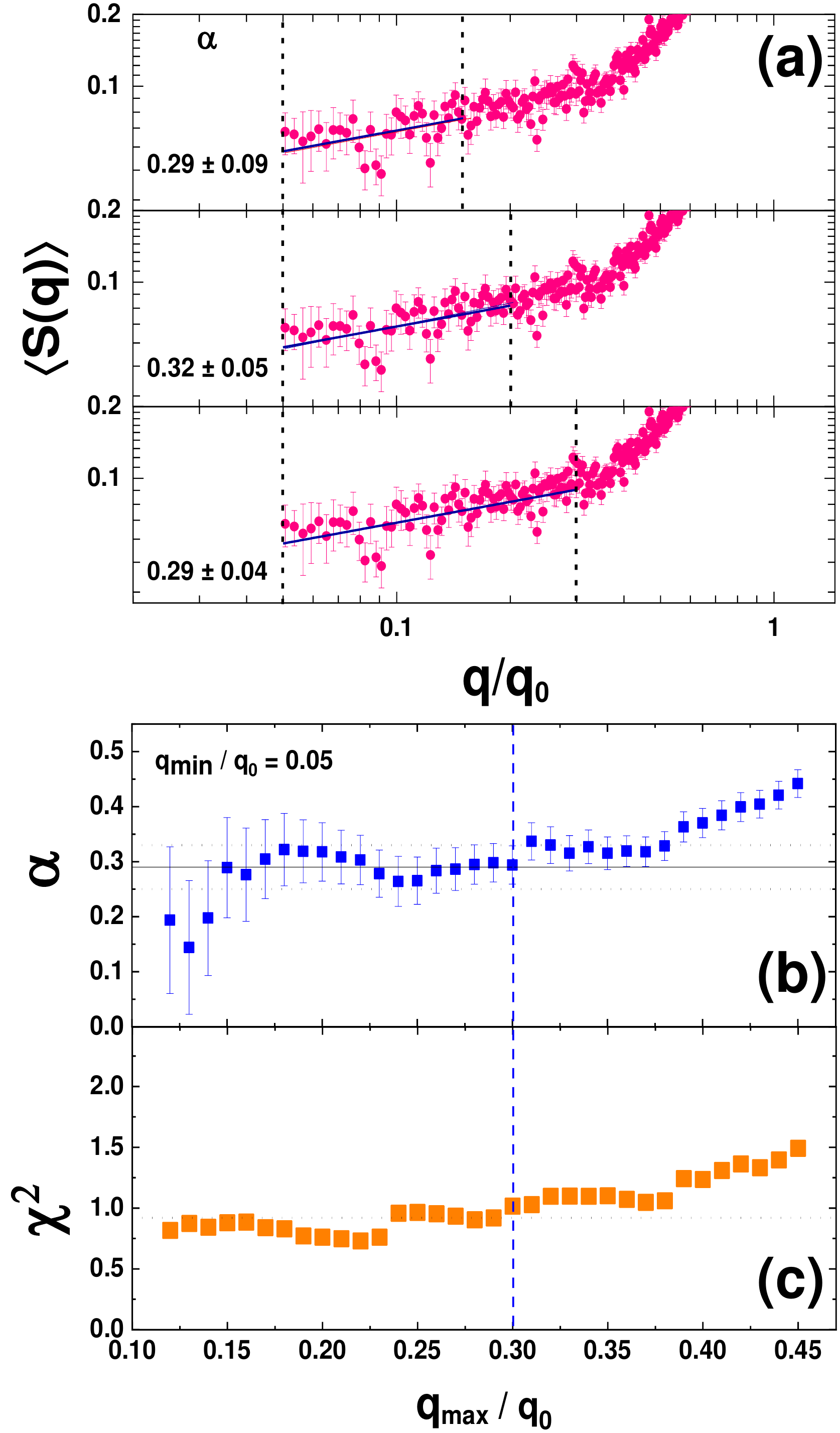}
    \caption{(a) Algebraic-growth exponents obtained from fitting the angularly-averaged structure factor data in three different low-$q$ ranges indicated by vertical black dotted lines. The fitted exponents are similar within the error. The data presented in Fig.\,\ref{fig:3} (a) corresponds to the fitting range of the bottom panel with $q/q_{\rm 0}$ ranging between 0.05 and 0.3. (a) Fitted algebraic growth exponent $\alpha$ as a function of the fitting range, extending from the lowest accessible wavevector $q_{\rm min}/q_{\rm 0}=0.05$ up to a variable $q_{\rm max}/q_{\rm 0}$. The vertical blue dashed line indicates the $q_{\rm max}/q_{\rm 0}=0.3$ considered to obtain the fit of $\langle S(q) \rangle$ data shown in Fig.\,\ref{fig:3} (a). The horizontal black line corresponds to $\alpha = 0.29$. (b) $\chi^{2}$ goodness of the fit for different fitting ranges. }
     \label{fig:14}
\end{figure}

In order to provide evidence on the robustness of the fitted algebraic exponent $\alpha$=$0.29 \pm 0.04$,
 Fig.\,\ref{fig:14} (a) shows that this value is, within our error estimation, stable under changes of the fitting region. This is more systematically quantified in Fig.\,\ref{fig:14} (b) where we plot the fitted algebraic growth exponent $\alpha$ as a function of the fitting range, extending from the lowest accessible wavevector $q_{\rm min}/q_{\rm 0}=0.05$ up to a variable $q_{\rm max}/q_{\rm 0}$. This figure shows that there is a plateau or range of  $q_{\rm max}/q_{\rm 0}$ where $\alpha$  is well defined, fluctuating within the error around the  value of $0.29 \pm 0.04$   (see black horizontal line). A departure from the plateau is observed at very small  $q_{\rm max}/q_{\rm 0}$ since data is very noisy and is affected by the finite size of the experimental field of view. A departure from the plateau is also observed at large  $q_{\rm max}/q_{\rm 0}$ due to the crossover to a different regime in the structure factor when the Bragg wavevector $q_{\rm 0}$  is approached. In order to further support the robustness of the value of the growth exponent of $\langle S(\bf q) \rangle$, Fig.\,\ref{fig:14} (c) shows that the $\chi^{2}$ goodness of the fit is quite acceptable precisely in the plateau where $\alpha = 0.29 \pm 0.04$. Thus, this analysis strongly supports that the long-wavelength vortex density fluctuations follow a class-III hyperuniform behavior.

In addittion,  the vortex system in FeSe is \textit{isotropically} class-III hyperuniform despite the anisotropic interaction term due to the magneto-elastic effect.
This is evident in the angularly-disaggregated data shown in the insert of Fig.\,4 (a) where we plot $\langle S_{\rm i} ({\bf q}) \rangle $, the averaged structure factor in each of the six angular sectors highlighted in the two-dimensional structure factor pattern. The main point of this figure is to pass two messages: (1) The $\alpha = 0.29 \pm 0.04$ obtained from the fit of $\langle S(\bf q) \rangle$, see black line in the figure, is consistent within the error with values obtained from angularly-disaggregated data; (2) $S(q_{\rm x}, q_{\rm y})$ vanishes isotropically in the $q \rightarrow 0$ limit, thus bringing evidence of disordered hyperuniformity with $S(q_{\rm x}, q_{\rm y}) \equiv{S(\bf q)} \sim q^{0.29}$.

\begin{figure}[ttt]
    \centering
    \includegraphics[width=\columnwidth]{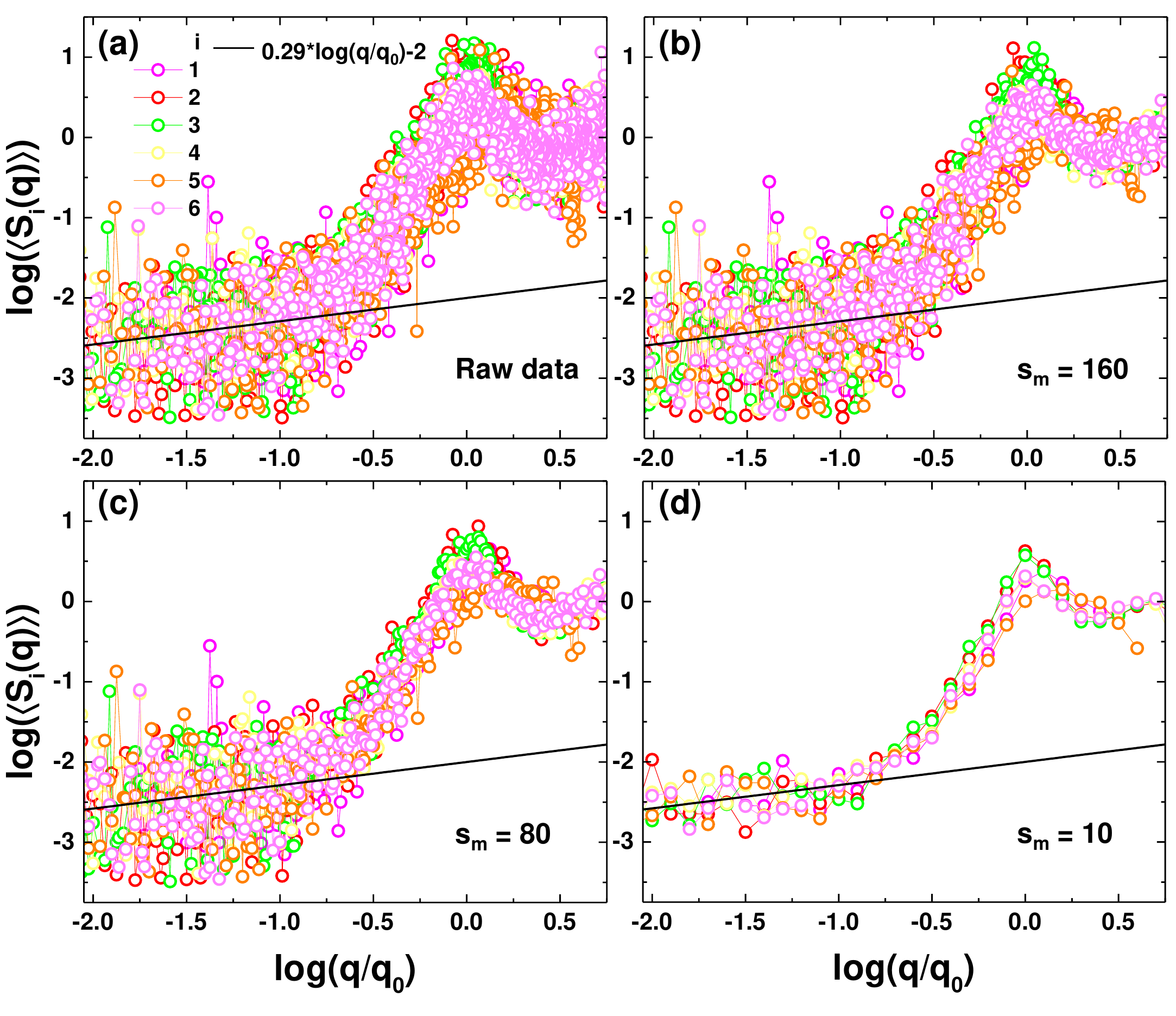}
    \caption{Smoothed angularly-disaggregated structure factor data $\langle S_{\rm i} ({\bf q}) \rangle $ data in the 6 angular sectors indicated in the insert of Fig.\,3 (a) of the main text. To smooth,$\langle S_{\rm i} ({\bf q}) \rangle $ data with an identical  $[\log(q/q_0)s_{\rm  m}]/s_{\rm  m}$ value, where $[...]$ is the integer part operator and $s_{\rm  m}$ is a smoothing parameter (the smaller $s_{\rm  m}$  the larger the logarithmic bins). We show data in the panels (a) the raw data, and different smoothing parameters of (b) 160, (c) 80 and (d) 10.  The black line is an algebraic fit to the data yielding a value $\alpha = 0.29 \pm 0.09$.}
     \label{fig:12}
\end{figure}

Although  the experimental disaggregated data have large fluctuations in the small-$q$ region, these two messages can be clearly extracted from the results of a progressive smoothing over $q$ of the $\langle S_{\rm i} ({\bf q}) \rangle $ data. To smooth the data we used a logarithmic binning in order to avoid a distortion of the power law behavior.  In order to smooth the data, we average $\langle S_{\rm i} ({\bf q}) \rangle $ data with an identical  $[\log(q/q_0)s_{\rm  m}]/s_{\rm  m}$ value, where $[...]$ is the integer part and $s_{\rm  m}$ is a smoothing parameter (the smaller $s_{\rm  m}$  the larger the logarithmic bins). Figure\,\ref{fig:12} shows the result of smoothing for different values of $s_{\rm  m}$ and the black line corresponds to a fit to the data yielding $\alpha = 0.29 \pm 0.09$, the same exponent found in the case of the angularly-averaged $\langle S(\bf q) \rangle$ data within the error. This brings further evidence that at small $q$ the amplitude of vortex density fluctuations vanishes isotropically since the same smoothing is applied to every direction.

Finally, it is important to recall that, despite the vortex structure in FeSe is disordered hyperuniform, the decay of $G_{6}$ with distance indicates that there is a non-negligible locking of the vortex structure to a particular crystal direction. According to $S(\mathbf{q})$ and Delaunay triangulation data this direction corresponds to  55\,$^{\circ}$ from the $\mathbf{b}$-axis.

\subsection{Simulated vortex structures with an extra anisotropic long-range interaction term}

	We now study the possibility of the coupling with the host medium inducing the weakest class-III hyperuniformity in a material with point disorder that would otherwise present class-II hyperuniformity.~\cite{Rumi2019}  First, we argue that for low $B$ the electronic nematicity of FeSe seems not to play a determinant role in the vortex structure. The round-like magnetic halo of vortices imaged at 5\,G (see Appendix E) differs from the nematicity-induced elliptical vortex shape observed at large fields.~\cite{Putilov2019,Hanaguri2019}  Nematicity might induce a transition to an oblique vortex structure, but the low field phase is an undistorted hexagon.~\cite{Lu2018} Second,  short-range non-local electronic coupling is expected to induce this transition, but the low-field phase would not present rhombic distortions neither.~\cite{Kogan1997} This effect is dominant for low-$\kappa$ weak coupling superconductors, conditions hardly fulfilled in FeSe ($\kappa$$=$$55$ and  $2\Delta/k_{\rm B}T_{\rm c}$$\sim$$6.3$). Third, the long-range anisotropic magneto-elastic effect
	arising from the weak elastic crystal perturbations induced by the nucleation of vortices has a magnitude $\eta \propto (dT_{\rm c}/dP)^{2}$. In FeSe   $dT_{\rm c}/dP$$\gtrsim$$10$\,K/GPa,~\cite{Margadonna2008} hundred times larger than in conventional superconductors. Thus, the magneto-elastic effect is the dominant coupling of the vortex system with the FeSe host medium.

	Here we show that the long-range anisotropic vortex interaction term arising from magneto-elasticity, besides producing rhombic distortions and orientational locking, induces the weakest class-III hyperuniformity. We perform two-dimensional finite-temperature Langevin dynamics simulations following the technical details described in the Experimental section. In order to mimic the experimental field-cooling, we start simulations from a highly disordered configuration and let the system relax at an effective temperature  per unit length ${\cal T}_{\rm eff}$ until the total interaction energy reaches a plateau. With the aim of reproducing the experimental data, we compute $\langle S({\bf q}) \rangle$ for final configurations varying  $\eta$ and ${\cal T}_{\rm eff}$.

	Figure\,\ref{fig:3} (b) shows the results of the simulation with the best quantitative fit with the experimental data ($\eta$$=$$0.1$ and $k_{\rm B}{\cal T}_{\rm eff}$$=$$0.004\epsilon_{0}$).
 The right panel shows a zoom into a snapshot of a typical simulated structure. Many vortex rows are at $\pm 45$\,$^{\circ}$ from $\mathbf{b}$, the directions where the magneto-elastic interaction is attractive.

 Details on how the changes on the effective temperature and
 strength of the magneto-elastic coupling produces a departure from the structural properties observed experimentally are presented in Appendix F.  This optimal value of $\eta$ is roughly the expected  for a material with  $dT_{\rm c}/dP \gtrsim 10$\,K/GPa.~\cite{Lin2017}
  If this magnitude is decreased to half, $\eta$ decreases by a factor of four $\sim 0.2$ and in that case there is no agreement between the simulated and the experimental structures. Thus, we estimate that if the host media have a $dT_{\rm c}/dP \sim 0-5$\,K/GPa no important spoiling of the hyperuniformity class of vortex matter is expected. Therefore,  the hyperuniformity class of vortex matter can be tailored by adequately choosing the superconducting sample as to tune the value of $\eta$ in the mentioned range of values.

 The data obtained from averaging over tens of configurations following  a similar field cooling protocol as in the experiments show a very good agreement with the experimental results. For instance, Figs.\,\ref{fig:3} (b) and (c) show typical simulation and experimental results in regions with the same number of vortices. Non-sixfold coordinated vortices are highlighted in red whereas sixfold coordinated are shown in blue. These typical images illustrate what is observed in the larger field of view structures: The experimental and simulated patterns have a similar value of topological defects,  size of crystalites, and present the same type of large-scale density fluctuations as observed in the angularly-averaged $S(q)$ data of panel (a) of Fig.\,\ref{fig:3}. Both structures also present quasi long range orientational order, but  the overall orientations of the simulated (experimental) vortex structure is at 45\,$^{\circ}$ (55\,$^{\circ}$) from the $\mathbf{b}$-axis.  This is connected to the fact that the structure factor for the simulated pattern displays diffraction spots at the vertices of a square instead at those of a rectangle as observed in the experimental data. This difference is only relevant for the structural properties of vortex matter at wavevectors $q \sim q_{0}$.  Nevertheless, in the $q \rightarrow 0$ limit the $\langle S({\bf q}) \rangle$ obtained in simulations is isotropic and coincides with the experimental data, see violet squares in Fig.\,\ref{fig:3} (a).

 Finally, we would like to recall that these simulation results are not altered by explicitly including  the effect of weak random pinning that significantly increase the computational time. We show in Appendix G that the effect of pinning is captured in our simulations by considering a system without pinning but with a slower dynamics.

\section{Discussion}

\begin{figure}[ttt]
    \centering
    \includegraphics[width=1\columnwidth]{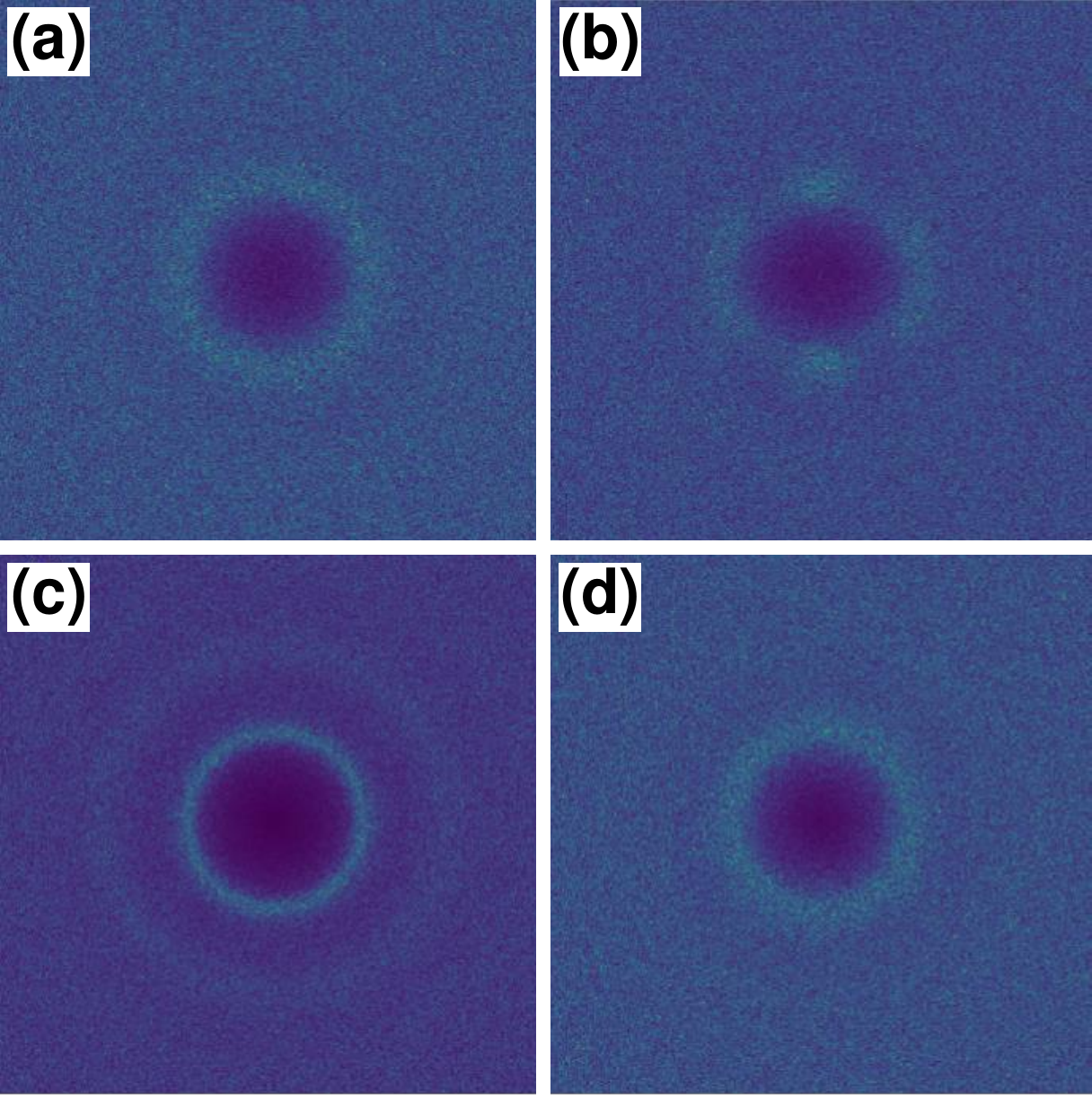}
    \caption{Structure factor of simulated vortex structures considering different vortex-vortex interaction models: (a) London model $U(r)\sim K_0(r/\lambda)$; (b) Anisotropically modulated London model with $U\sim K_0(r/\lambda)(1+\cos(4\phi)/2)$; (c) Coulomb $U\sim 1/r$ repulsion; (d) $U\sim 1/r^2$ repulsion. In all simulations we consider $8192$ vortices with an average spacing $a_{0} = 6.5 \lambda$ and $k_{\rm B}{\cal T}_{\rm eff}=0.004 \epsilon_{0}$.}
    \label{fig:10}
\end{figure}

In order to provide stronger support for our claim that the
magneto-elastic effect is responsible for degrading the hyperuniformity of the system to class-III, now we discuss a qualitative comparison between the structural properties of vortex systems obtained in simulations considering different interaction models and the experimental results in FeSe.
 To start with, we consider the standard case in most three-dimensional superconductors of an isotropic London interaction between vortices ($\eta=0$ in Eq.\ref{eq:koganmodelforces}).
Figure\,\ref{fig:10}\,(a) shows the structure factor obtained in this case for the same vortex density and effective temperature as considered in our former simulations, see Section II B for details.  The diffraction pattern is ring-like indicating that the system presents a liquid-like structure at this density. This type of order is observed experimentally in vortex structures nucleated at low fields in materials with a negligible magneto-elastic effect.~\cite{AragonSanchez2019}  Thus, anisotropy in the interaction term is required to induce non-isotropic distortions in the vortex structure as the rhombic ones observed experimentally in FeSe.

Nevertheless, anisotropy alone does not produce vortex structures with the structural properties observed experimentally. For example, now we consider
 anisotropic interactions with the same quadrupolar symmetry than the magneto-elastic but short-ranged, described  by the anisotropically modulated London model
\begin{eqnarray}
U(r,\theta)/\epsilon_0= K_0(r/\lambda)( 1+ {\tilde \eta} \cos(4\theta)).
\end{eqnarray}
\noindent This is a particular case of the more general interaction form used in Ref.\,\onlinecite{Olszewski2018}, with $\tilde{\eta}$ controlling the strength of the anisotropic contribution. Figure\,\ref{fig:10}\,(b) shows the $S(\mathbf{q})$ data considering this interaction that is closer to the experimental results. These data are obtained for  $\tilde{\eta}=0.5$, much larger than the $\eta=0.1$ used to reproduce the experimental data using the Kogan interaction (data of Fig.\ref{fig:8}). The structure factor in the case of this anisotropically modulated London model presents some diffuse peaks but their angular location in $q$-space differs from the peaks observed experimentally in FeSe.  Then both, long-ranged and anisotropic interactions in register with the crystal structure are mandatory to satisfactorily describe the structural properties of vortex matter nucleated in FeSe.

In order to strengthen this last assertion,
Figs.\,\ref{fig:10}\,(c) and (d) show the structure factors obtained in simulations when considering other types of isotropic but long-ranged interactions between vortices. Panel (c) depicts the result obtained in the case of a repulsive $ U \sim 1/r$ Coulomb interaction whereas panel (d) corresponds to a purely repulsive $U \sim 1/r^2$ interaction. In both cases the structure is isotropic though the local order is, as expected, stronger for the case of Coulomb interactions. Coulomb-like interactions have been argued to be effective two-dimensional interactions in the top layer of three-dimensional objects if transverse fluctuations along the $\mathbf{c}$-direction are integrated out, while the $1/r^2$ interaction is typical from elastic interaction kernels.

Therefore, only a vortex-vortex interaction that is both, long-ranged and anisotropically locked to the sample crystal structure, can reproduce the observed FeSe vortex structure presenting rhombic distortions and class-III hyperuniformity at low vortex densities. This kind of
interaction is compatible with the  magneto-elastic coupling between the vortices and the host superconducting crystal.

\section{Conclusions}	
	
	In conclusion, the magneto-elastic interaction plays a determinant role in weakening hyperuniformity in diluted vortex structures. This is an example of the relevance of the coupling between the interacting objects and the host medium on determining the hyperuniformity-class of the former.  Whether this is the only mechanism to induce class-III hyperuniformity remains an interesting open question. This work opens a new avenue on how to tailor the hyperuniformity-class of material systems  by means of tuning the  elastic and electronic properties of the host media where interacting objects are nucleated.
	
\section{Acknowledgments}	
	
	Work funded by the Ministry of Science (grants PICT 2017-2182, 2018-1533, 2019-01991) and Cuyo National University (grants 06/C566 and
	06/C575) of Argentina. Y. F. thanks  the Alexander von Humboldt Foundation for founding through the Georg Forster Research Award.

\section*{Appendix A:Sample growth and characterization}

\begin{figure}[bbb]
\begin{center}
\includegraphics[width=\columnwidth]{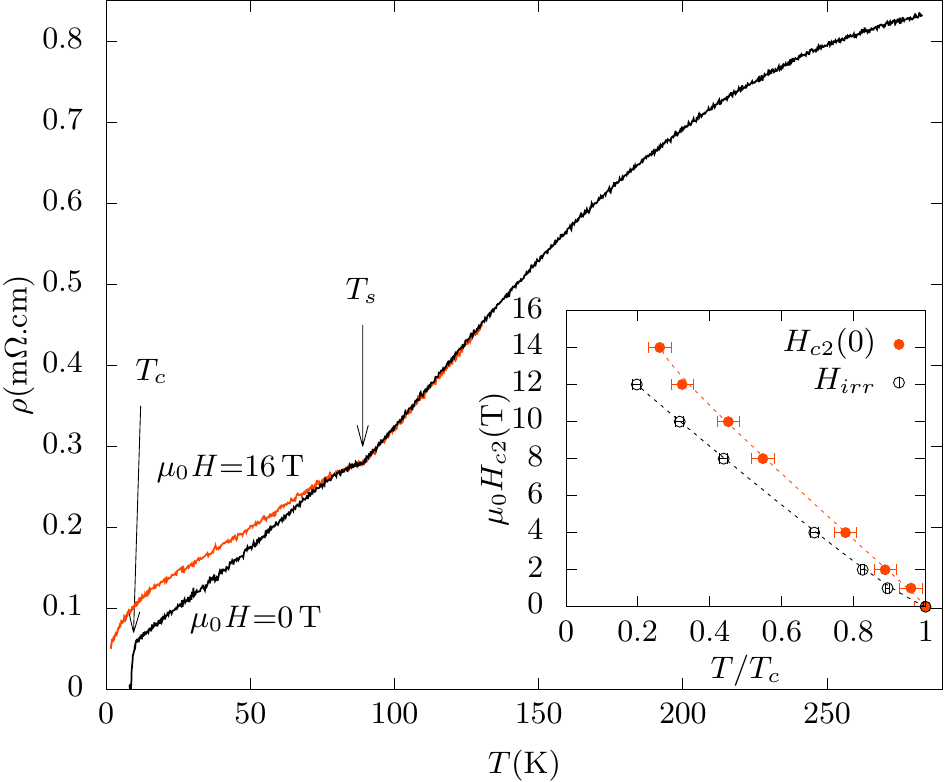}
\end{center}
\caption{Temperature dependence of the $ab$-plane resistivity of $\beta$-FeSe for  magnetic fields applied along the $\mathbf{c}$-axis of 0 and 16\,T. The temperatures of the structural, $T_{\rm s}$, and superconducting transitions, $T_{\rm c}$, are indicated with arrows. Insert: Superconducting critical field, $H_{\rm c2}$, and irreversibility field, $H_{\rm irr}$, as a function of the reduced temperature.}\label{rvst}
\end{figure}

We grow single crystals of $\beta$-FeSe  applying the vapor transport method using AlCl$_3$/KCl as flux.~\cite{Amigo2014}  We obtained platelet-shaped crystals with the tetragonal $\beta$-FeSe phase only, as checked by X-ray diffraction measurements. We performed transport measurements in the four-probe configuration varying temperature at fixed fields of up to 16\,T. Figure \ref{rvst} presents $ab$-plane resistivity results for a crystal of the same batch than the ones studied by magnetic decoration.
These crystals present two characteristic temperatures: The structural transition from tetragonal to orthorhombic symmetry at $T_{\rm s} \sim$ 90\,K, and the   superconducting transition temperature $T_c$=9.6(2)\,K.
The increase of the resistivity in the normal state with increasing magnetic field is a characteristic of the multiband nature of FeSe. \cite{Amigo2014}
The inset of Fig.\,\ref{rvst} shows the superconducting critical field, $H_{\rm c2}$, and the irreversibility field, $H_{\rm irr}$.

\section*{Appendix B: Data in a smaller sample with no twins detected}

\begin{figure}[bbb]
    \centering
    \includegraphics[width=\columnwidth]{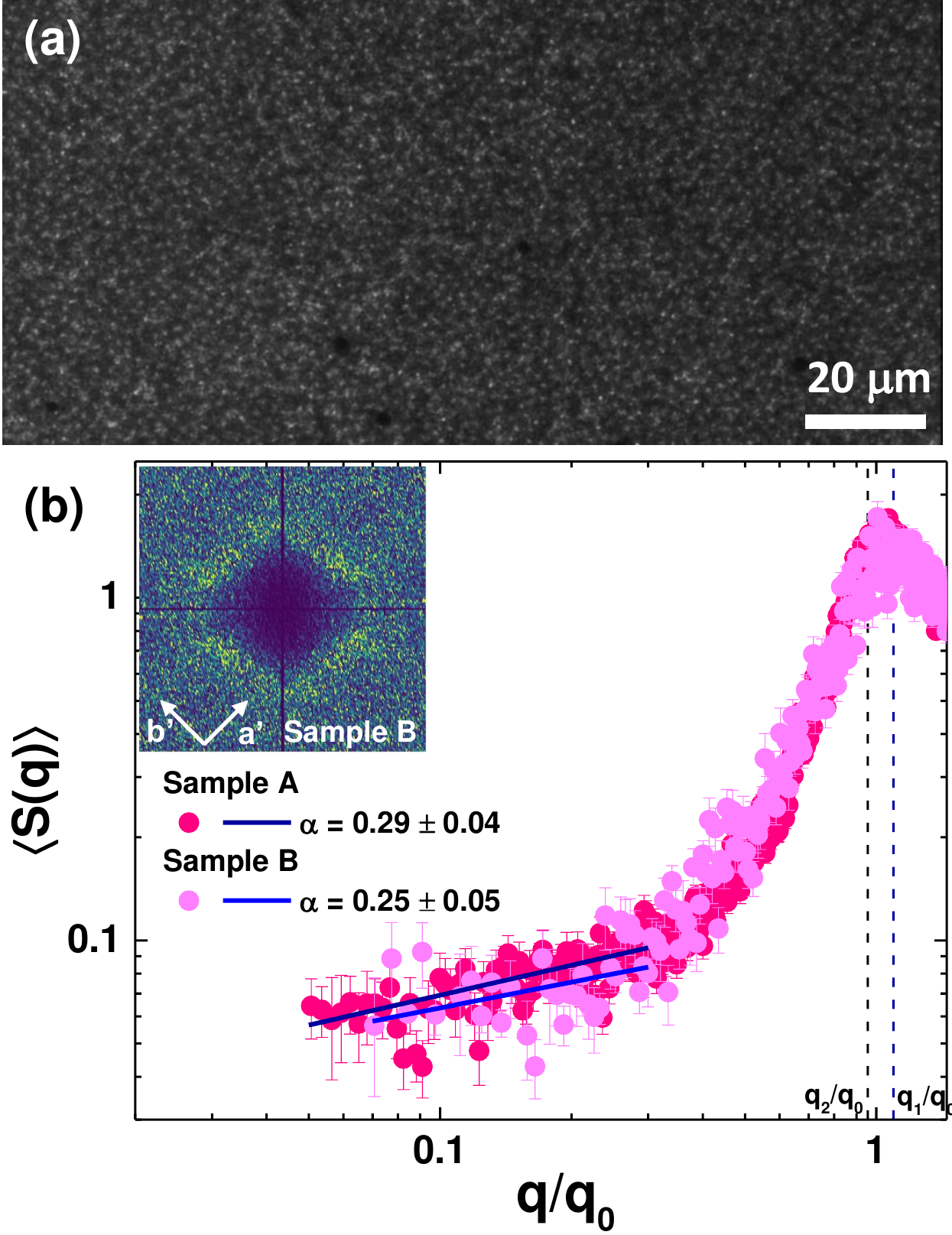}
    \caption{(a) Magnetic decoration image of vortex matter nucleated at 5\,G in an untwinned FeSe crystal B. This sample is smaller than the sample A presented in the main text: Only 2500 vortices where imaged in contrast to 13586 vortices imaged in
    sample A. (b) Angularly-averaged structure factor for samples A and B and corresponding algebraic fits of the data with exponents $\alpha$ similar within the error bars (see legend). Insert: Structure factor of the vortex structure in sample B with the crystal unit cell vectors indicated.}
    \label{fig:fig13}
\end{figure}

The results presented in the main text correspond to data in sample A where 13586 vortices are imaged and no twin boundaries are detected in the whole sample. In another smaller sample from the same batch, sample B, also no twin boundaries were detected but only 2500 vortices are imaged in this sample, see  Fig.\,\ref{fig:fig13} (a). In the case of the vortex structure nucleated in sample B,  Fig.\,\ref{fig:fig13} (b) shows the structure factor and its angular average $\langle S(\bf q) \rangle$. The same figure presents data of sample A for comparison: The $\langle S(\bf q) \rangle$ data are similar in both samples within the error. Since a smaller amount of vortices is imaged in sample B than A, the minimum modulus of the wavevector down to where we can compute the structure factor is larger in sample B. This reduces the fitting range of $\langle S(\bf q) \rangle$ in the case of sample B. In the case of data in sample A presented in the main text the fitting range is between $q/q_{\rm 0}=0.05$ and 0.3.  Nevertheless, if both sets of data are fitted with an algebraic growth in the range between the minimum detected wavevector and $q/q_{\rm 0}=0.3$, similar $\alpha$ exponents are found within the error of the fit, see details in main panel of Fig.\,\ref{fig:fig13} (b).

\section*{Appendix C: Twin boundaries in some FeSe samples as imaged by magnetic decoration}

\begin{figure}[bbb]
    \centering
    \includegraphics[width=\columnwidth]{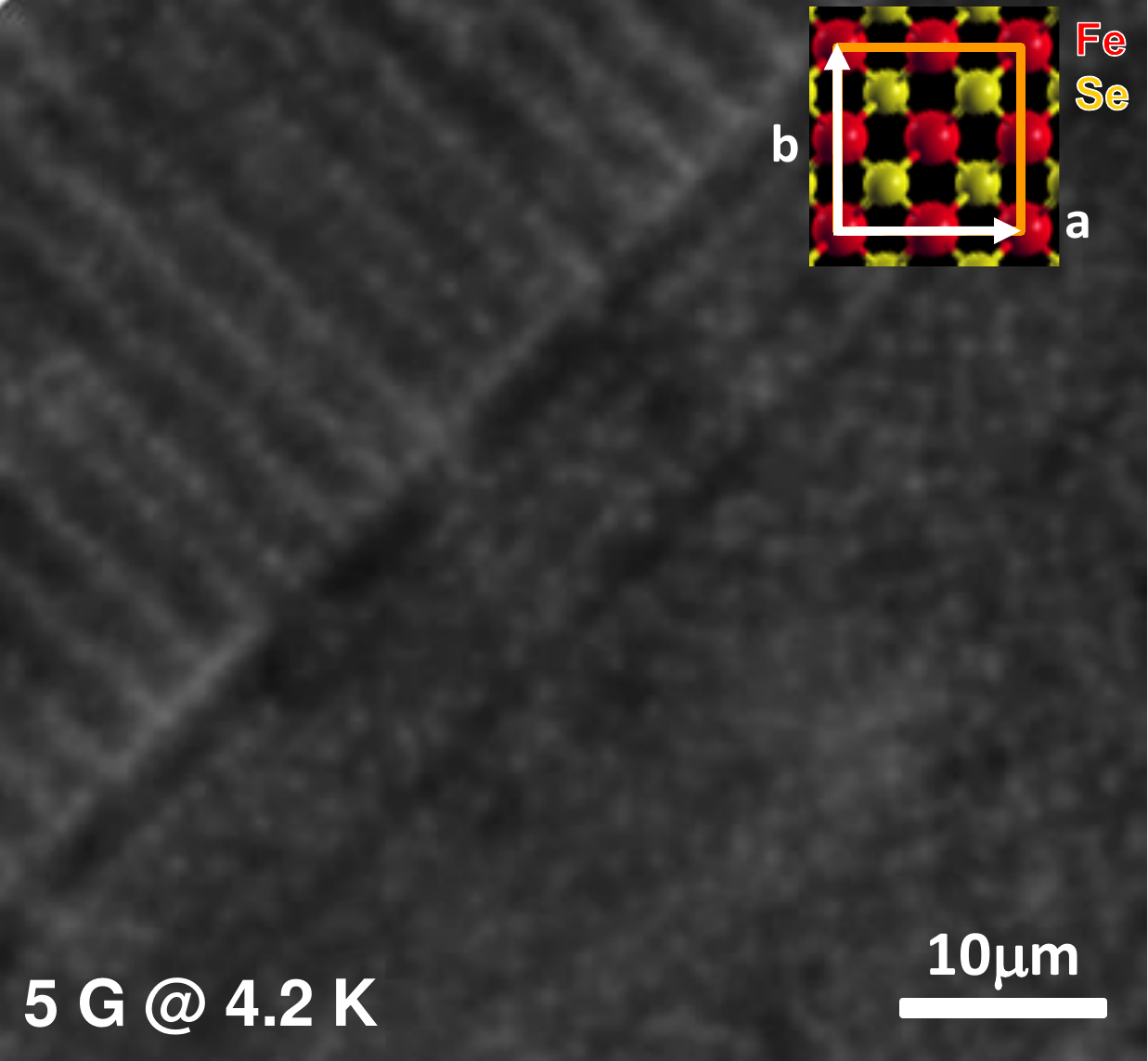}
    \caption{Magnetic decoration image of vortex matter nucleated at 5\,G in a FeSe crystal presenting several correlated defects. Insert: Orientation of the orthorhombic crystal structure in this sample, though whether $a$ is the shortest or largest unit cell vector can not be ascertained since X-ray measurements were performed in the tetragonal phase.}
    \label{fig:SMdeco}
\end{figure}

Figure\,\ref{fig:SMdeco} shows a snapshot of the vortex structure nucleated after a field cooling process at 5\,G. The magnetic decoration experiment was performed at 4.2\,K.  In this sample, Fe-Fe bond directions indicated as the crystallographic axes $\vec{a}$ and $\vec{b}$ are aligned with the borders of the picture, see white arrows in the schematic drawing of the insert in Fig.\,\ref{fig:SMdeco}. Whether $a$ is the shortest or largest unit cell vector can not be ascertained since X-ray measurements were performed at room temperature, in the tetragonal phase of the material. The observed correlated defects are aligned at $45^{\circ}$  from the $a$ and $b$ axes, namely in the Fe-Se bond directions. In most of the studied 30 samples from the same batch,  twin boundaries were detected. Only in two samples, samples A and B, no twin boundaries were detected in the whole sample.

\section*{Appendix D: Increase  of the variance of the number of vortices with distance}

\begin{figure}[bbb]
    \centering
    \vspace{2cm}
    \includegraphics[width=1.05\columnwidth]{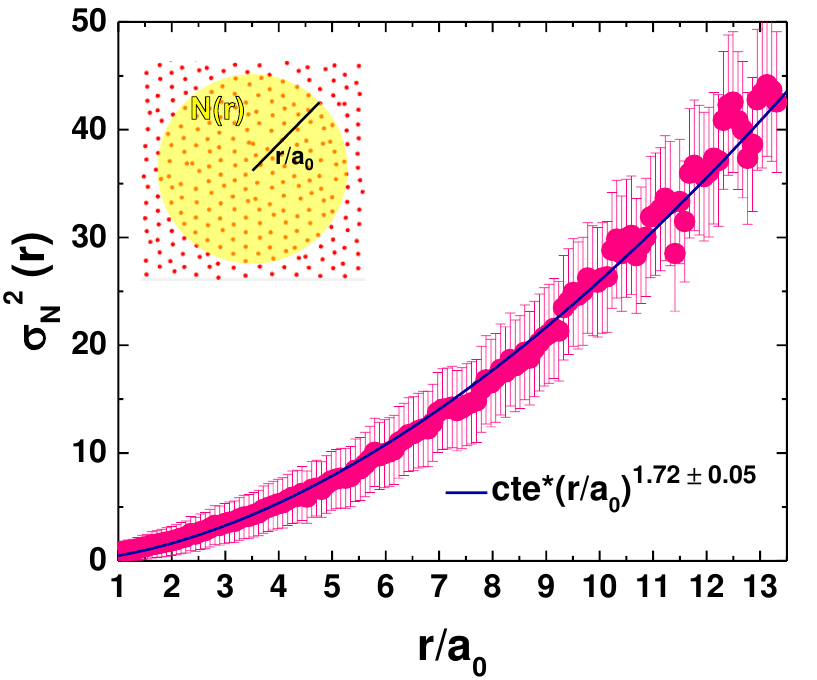}
    \caption{Increase of the variance $\rm \sigma^{2}_{\rm N}$ of the vortex number $N(r)$ with $r/a_{0}$:  Algebraic growth yielding an exponent of $1.72 \pm 0.05$ (blue line). Error bars correspond to the standard deviation of data obtained for circles with the same $r/a_{0}$. Insert: Computation of the number of vortices   in circular windows with radius $r/a_{0}$.
    }
     \label{fig:6}
\end{figure}

The disordered  hyperuniformity of the low field FeSe vortex structure observed from the decay of the structure factor in the $q \rightarrow 0$ limit is also quantitatively supported by the growth with distance $r$ of the vortices number variance, $\rm \sigma_{N}^{2} (r) = \langle N^2(r)\rangle - \langle N(r)\rangle^{2}$. The  insert in Fig.\,\ref{fig:6} shows a schematics of the calculation of this magnitude:  The number of vortices $\rm N(r)$ inside an area $\rm \pi (r/a_0)^2$ changes for different values of $\rm r/a_0$ and also depends on the location of the circular area.  We calculated $\rm \sigma_{N}^{2}(r)$ over thousands of circles centered at random in the whole field of view. In the case of data for sample A shown in the main text, this magnitude grows with distance with an exponent $\beta = 1.72 \pm 0.05$. It should be noted that for perfectly rhombic or hexagonal structures the expected exponent is $\beta=1$ while for random structures $\beta=2$. The values of the exponents $\alpha$ and $\beta$ found experimentally support the low-field vortex structure in FeSe is class-III hyperuniform since theoretically a relation $\beta = 2 - \alpha$ is expected~\cite{Torquato2018} for a two-dimensional pattern of points such as the tips of vortices we are observing at the sample surface.

\section*{Appendix E: Magnetic halo of vortices at low fields}

Previous STM works reveal that at high magnetic fields the spectroscopic halo of vortices is elliptical.~\cite{Song2011,Chowdhury2011,Watashige2015,Hanaguri2019,Putilov2019} This is interpreted as the consequence of a strong electronic nematicity manifested as an elliptic multi-band Fermi surface. FeSe has an electron band whose minor axis is oriented along the $a$ direction and a hole band with its minor axis oriented along the $\mathbf{b}$-axis. In these works these unit vectors are defined considering that in the orthorhombic low-temperature ($T \lesssim 90$\,K) phase $a<b$. For elliptical Fermi surfaces, in the vortex core the number of quasiparticles (excitations of the superconducting state) that travel along the minor axis is larger than the number of quasiparticles travelling in the largest axis. Thus, the vortex core, or the physical place where quasiparticles live, develops an elongated shape along the shortest axis of the Fermi surface. STM results at high fields indicate vortices are elongated along the $b$ direction, and thus the elongated shape of vortices is associated to the hole band of the multi-band Fermi surface in FeSe. The magnitude of the eccentricity of the elliptical vortices can change with field, and indeed in Ref.\,\onlinecite{Putilov2019} it seems to decrease its magnitude for smaller vortex densities and the orientation of the shortest axis of the ellipse has local deviations from the $b$ direction.

\begin{figure}[ttt]
    \centering
    \includegraphics[width=\columnwidth]{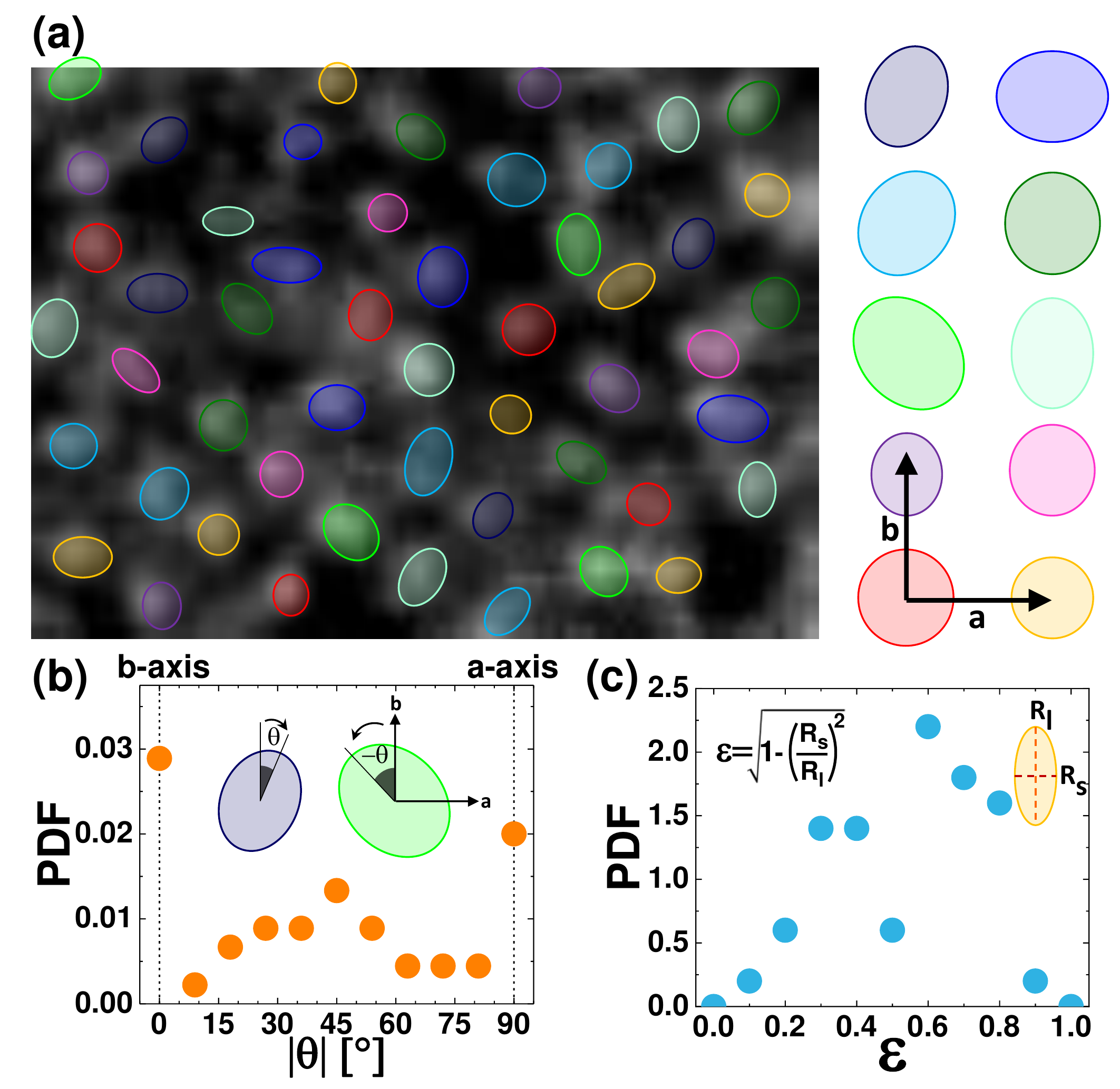}
    \caption{Eccentricity and orientation of the magnetic halo of vortices in FeSe for diluted vortex densities of 5\,G. Data are obtained in sample A presented in the main text. (a) Zoom-in of a magnetic decoration image where the magnetic halo of individual vortices is fitted with an elliptical shape (superimposed in various colours). Right: Schematics of the orientation of fitted ellipses with respect to the Fe-Fe bond directions $a$ and $b$ identified in the studied sample. (b) Probability density function of the orientation of the semi-major or long axis of the ellipses with respect to the $\mathbf{b}$-axis corresponding to $\theta=0$\,$^{\circ}$. (c) Probability density function of the eccentricity of the ellipses as defined in the schematics shown in the top part of the panel.}
    \label{fig:Maghalo}
\end{figure}

If the spectroscopic halo of vortices is elliptical, then the magnetic halo of vortices is expected to be elliptical. Thus, in order to ascertain  if the diluted vortex structures revealed in our study are strongly affected by electronic nematicity, we analyze the shape of the magnetic halo of vortices. Figure\,\ref{fig:Maghalo} (a) shows a zoom in to a magnetic decoration image where the shape of vortices was fitted with ellipses (coloured forms superimposed to the image). Fig.\,\ref{fig:Maghalo} (c) shows the definition of eccentricity $\epsilon = \sqrt{1 - (R_{\rm s}/R_{\rm l})^2}$ where $R_{\rm s}$ and $R_{\rm l}$ are the short and long axis of the ellipse, respectively. Thus $\epsilon=0$ represents the case of a vortex with a circular halo whereas $\epsilon \rightarrow 1$ corresponds to a very elongated vortex.  The probability density function of the eccentricity in the magnetic halo of vortices shown in Fig.\,\ref{fig:Maghalo} (c) follows a rather homogeneous distribution  in the 0.2-0.8 range, with no noticeable peak in this range. In contrast, a rough estimation  of the eccentricity of the spectroscopic halo of vortices nucleated in FeSe at 6\,T from the images shown in Ref.\,\onlinecite{Putilov2019} indicates its probability density function is peaked around $\epsilon \sim 0.95$ (corresponding to roughly a three times shorter minor than larger axis).

We also studied the orientation of the long or semi-major axis of the ellipse associated to the magnetic halo of vortices defining an orientation $\theta=0$\,$^{\circ}$ as the $b$ direction and $\theta=90$\,$^{\circ}$ as the $a$ direction. We recall that in this analysis we can not ascertain whether $a$ is the shorter or larger axis but we measured by XR diffraction that these are the two Fe-Fe bond directions.  At low vortex densities, the long axis is aligned along both crystallographic directions with roughly equal probability, see Fig.\,\ref{fig:Maghalo} (b). A lesser amount of vortices present their long axis aligned along intermediate directions. This situation is in strong contrast with the phenomenology observed at high fields in STM studies.~\cite{Putilov2019}

Thus, the moderate eccentricity of the magnetic halo of vortices and the non-preferential orientation of the large axis of the less-elongated vortices in one of the two Fe-Fe bonds directions indicate the effect of nematicity does not have a strong impact in the structural properties of vortex matter in FeSe for low vortex densities .

\section*{Appendix F: Impact of effective temperature and strength of magneto-elastic coupling on  simulations}

\begin{figure}[ttt]
    \centering
    \includegraphics[width=\columnwidth]{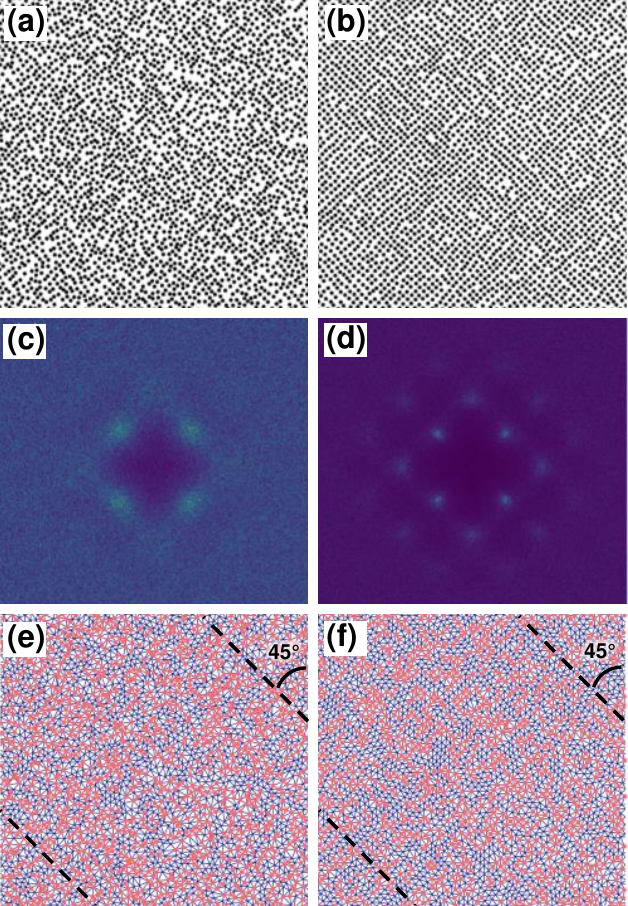}
    \caption{(a - b) Typical zoom-ins of snapshots, (c-d)  structure factors and (e-f) Delaunay triangulations of simulated vortex structures considering the London plus the magneto-elastic interaction terms. In both cases $\eta=0.1$ but the effective temperature is $k_{\rm B}{\cal T}_{\rm eff}=0.004 \epsilon_{0}$ for the left panels and $k_{\rm B}{\cal T}_{\rm eff}=0.001 \epsilon_{0}$ for the right ones.}
    \label{fig:8}
\end{figure}

\begin{figure}[ttt]
    \centering
    \includegraphics[width=0.98\columnwidth]{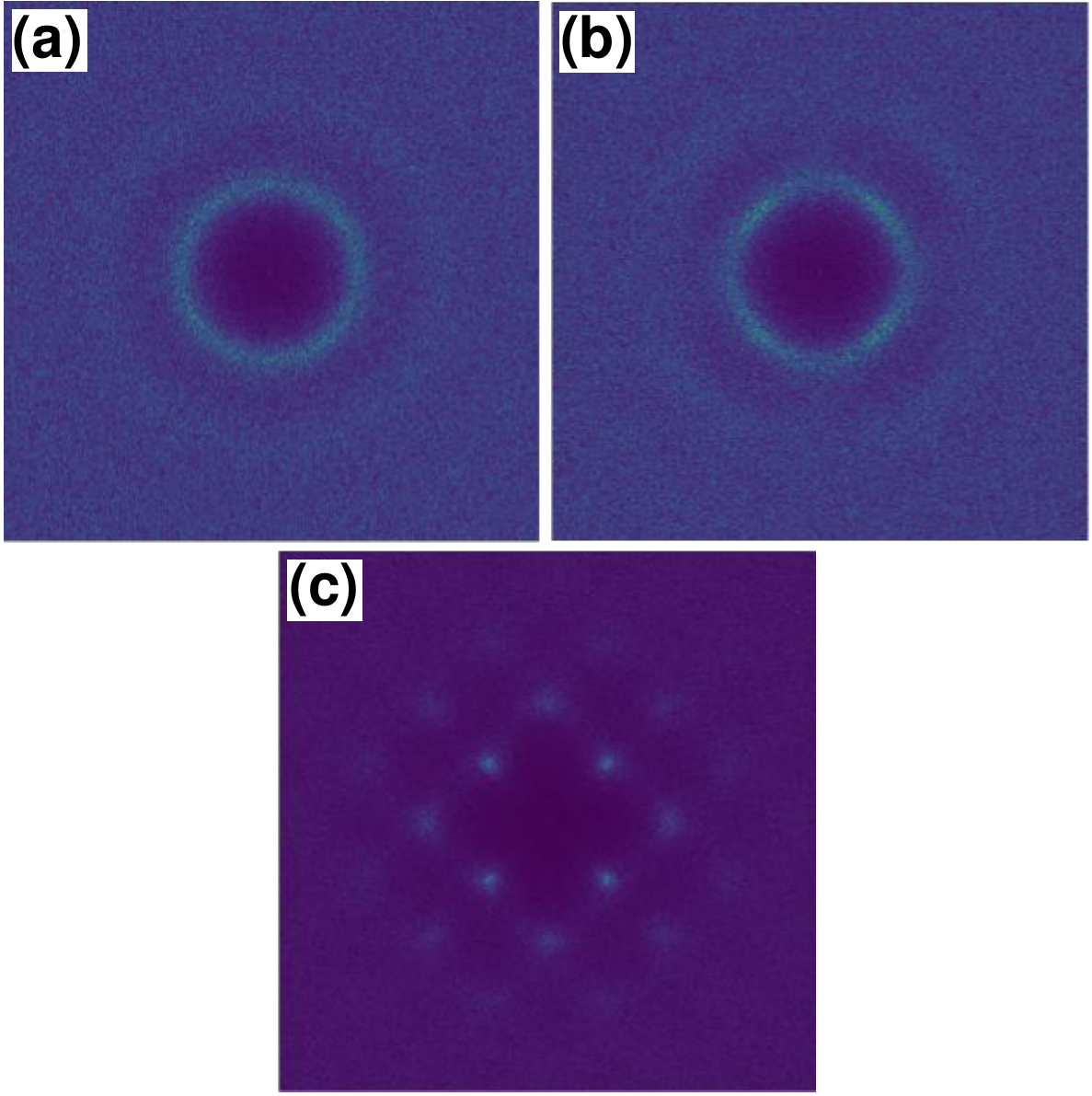}
    \caption{Structure factor of the vortex lattice considering London plus  magneto-elastic interaction terms for different magnitudes of coupling
    (a) $\eta= 0.001$, (b) 0.01 and (c) 0.1.  In all cases we performed simulations for an effective temperature $k_{\rm B}{\cal T}_{\rm eff}=0.001 \epsilon_{0}$.}
    \label{fig:9}
\end{figure}

We have found that the optimal values for qualitatively reproduce the experimental data are $\eta=0.1$ and $k_{\rm B}{\cal T}_{\rm eff}=0.004\epsilon_{0}$. Here we will show how sensitive the results are to deviations in both adjustable parameters.

In order to illustrate the effect of changing the temperature,
in Figs.\ref{fig:8} (a) and (b) we compare snapshots of configurations obtained for $\eta=0.1$ for two different effective temperatures: $k_{\rm B}{\cal T}_{\rm eff}=0.004\epsilon_{0}$ that is the optimal value for reproducing the experimental data and $k_{\rm B}{\cal T}_{\rm eff}=0.001\epsilon_{0}$.
Figures\,\ref{fig:8} (c) and (d) show their corresponding structure factors averaged over some hundreds of configurations.
Choosing a lower temperature allows us to diminish the entropy
and to observe the interaction energy-driven relaxation. We observe that both diffraction patterns present peaks at the same positions but become sharper and brighter on decreasing temperature. This indicates that the positional order enhances if the structure is quenched at lower temperatures. Nevertheless, irrespective of the temperature value, the structure presents a clear tendency to form a square lattice locked to the principal directions of the (assumed) tetragonal crystal structure ($x-y$ axis).

A close inspection of the zoom-ins of Figs.\,\ref{fig:8}(a) and (b) reveal that although the density of defects is smaller at lower temperatures, the structure resulting from simulations that do not explicitly consider quenched disorder still has a significant amount of topological defects. These defects are vacancies and disclinations, see the Delaunay triangulations of Figs.\ref{fig:8}(e) and (f). We also indicate the direction the displacement vector directin $\phi=\pi/4$ corresponding to one of the directions along which the magneto-elastic contribution in the pair interactions is purely attractive (the other is at $\phi=3\pi/4$, see Fig.\,\ref{fig:11}). As it can be better appreciated in Figs.\,\ref{fig:8}(a) and (b), vortices tend to form rows aligned along these two directions.

In order to illustrate the sensitivity with $\eta$ of our fit to experimental data, Fig.\,\ref{fig:9} compares the structure factors obtained by varying $\eta$ for a fixed $k_{\rm B}{\cal T}_{\rm eff}=0.001 \epsilon_{0}$.
The figure reveals that on decreasing $\eta$ one or two orders of magnitude than the optimal value of 0.1, the diffraction peaks are no longer defined and the system tends quickly to the isotropic order of a liquid phase.
It is worth recalling at this respect that the magnitude of the magneto-elastic coupling is predicted to be $\eta \propto (dT_{\rm c}/dP)^{2}$, thus relatively small changes in the slope of $dT_{\rm c}$ with pressure may translate into large changes of $\eta$.

\section*{Appendix G: Effect of pinning in our field-cooling vortex simulations}

\begin{figure}[ttt]
    \centering
    \includegraphics[width=0.76\columnwidth]{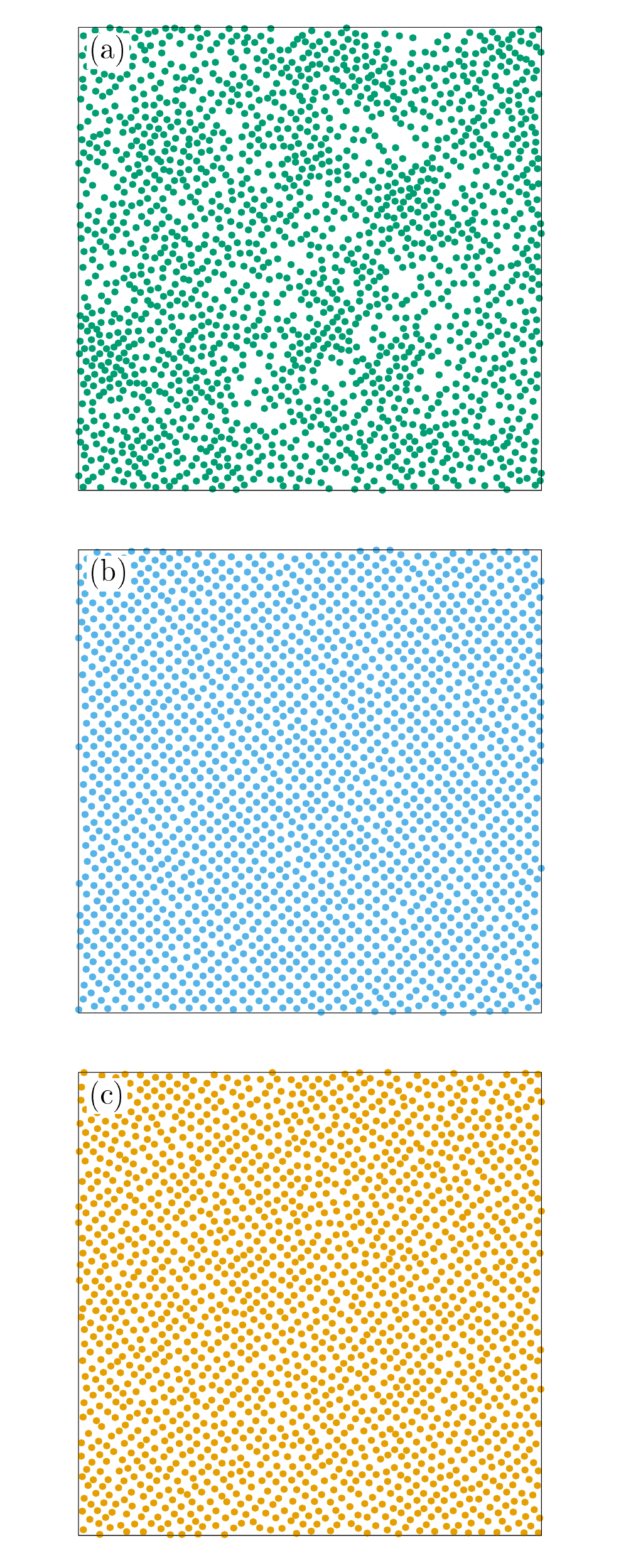}
    \caption{
    Zoom snapshots of particular configurations obtained from field-cooling Langevin simulations of a $N=8192$ particle system with effective magneto-elastic inter-vortex interactions. From a common initial condition (a), we show configurations at $t=768$ obtained without (b) and with a weak dense pinning (c).
    }
    \label{fig:PinVsSinPinSnap}
\end{figure}

\begin{figure}[ttt]
    \centering
    \includegraphics[width=0.99\columnwidth]{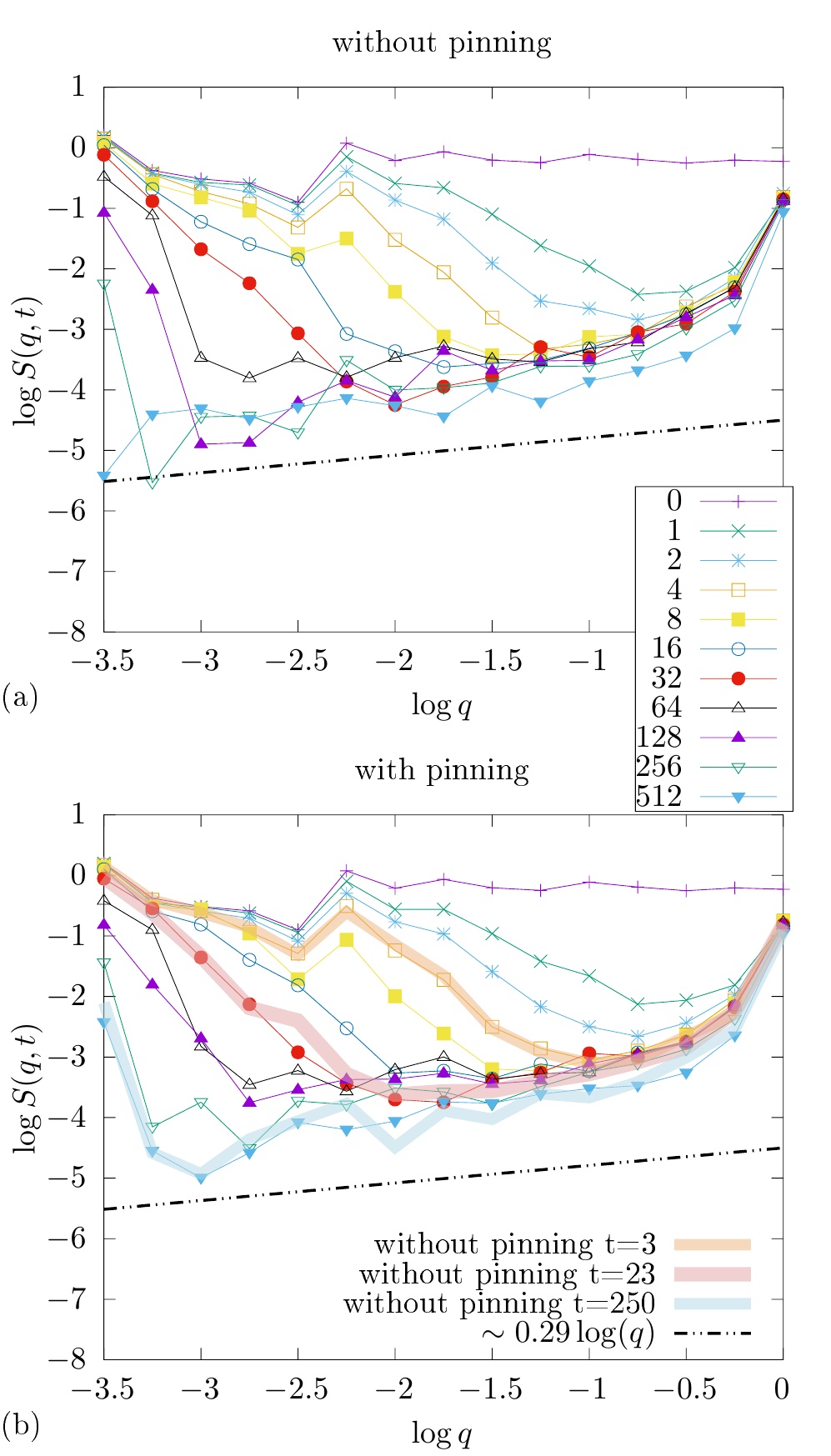}
    \caption{
    Time evolution of the smoothed structure factor obtained from Langevin dynamics simulation of $N=8192$ particles with mean separation $a_0\approx 3.0$, interacting with the effective magneto-elastic interaction potential, in the absence (a) and in the presence of pinning (b). The initial condition and cooling protocol are identical in both cases. In (b) the thick semi-transparent lines correspond to $S(q,t)$ obtained without pinning, but for different $t$, as indicated. In both panels, the thin dotted line indicates the hyperuniformity $S(q)\sim q^{-0.29}$ expected at longer times for small $q$.
    }
    \label{fig:PinVsSinPin}
\end{figure}

In order to include disorder we modify Eq.\ref{eq:langevin} as
\begin{eqnarray}
\alpha_{\rm BS} \frac{d{\bf r}_i}{dt}= \sum_{j\neq i} {\bf F}({\bf r}_i-{\bf r}_j)+{\bf F}_p({\bf r}_i)+\zeta_i(t).
\label{eq:langevinwithdis}
\end{eqnarray}
The quenched pinning forces are ${\bf F}_p({\bf r})=-\nabla U_{\rm p}({\bf r})$, where $U_{\rm p}({\bf r})$ is a random potential with adimensional correlation length $d=0.1$ and amplitude $A_p=0.1$. For the simulation we use $N=8192$ particles with a mean vortex spacing of $a_0\approx 3$. This value of $a_0$ is roughly half of the one used in the simulations presented in the main text. It was chosen so for practical reasons, to accelerate the relaxation in the case of pinning. The field-cooling protocol consists in decreasing the effective adimensional temperature as ${\cal T}(t)={\cal T}_0 + ({\cal T}_1-{\cal T}_0) t/t_{\rm ramp}$, with ${\cal T}_0=0.2$, ${\cal T}_1=0.0001$, $t$ the adimensional time and $t_{\rm ramp}=1000$ the adimensional ramp interval, being $\lambda \alpha_{BS}/ \epsilon_0$ the unit of time. During the process we monitor the angularly-averaged structure factor $S(q,t)$ vs $t$. The initial, highly-disordered state at $t=0$, has a Poisson distribution, and hence $S(q,t=0)\approx\text{const}$.

Two simulations were run following the same field-cooling protocol and identical initial conditions, one without pinning by solving Eq.(\ref{eq:langevin}), and the other with  dense weak point pinning, by solving Eq.(\ref{eq:langevinwithdis}).
In Fig. \ref{fig:PinVsSinPinSnap}(a) we show the configuration at  $t=1$, and in (b)-(c) the configurations at $t=768$ obtained without and with pinning, respectively.
Since we simulate only one disorder/thermal realization,
the $S(q,t)$ shown in Fig.\,\ref{fig:PinVsSinPin} was smoothed over $q$ in order to reduce the fluctuations. We used logarithmic binning in order to avoid a distortion of the emerging power-law at low $q$: It consists in averaging $S(q,t)$ over all $q$ with an identical $[\log(q) s_{\rm m}]/s_{\rm m}$ value, where $[\dots]$ denotes the integer part operator and $s_{\rm m}$ is a smoothing parameter ($s_{\rm m}=4$ in the data of Fig.\ref{fig:PinVsSinPin}).

The main results are shown in Fig.\ref{fig:PinVsSinPin} that shows  the evolution of $S(q,t)$ without (a) and with pinning (b). In both cases $S(q,t)$ evolves from an almost flat form $S(q,t=0) \approx \text{const}$ to an algebraically growing structure factor in the $q \to 0$ limit. It is evident from the plot that the whole evolution is approximately controlled by a growing correlation length scale $L(t)$. At a given time, modes such that $qL(t) \gg 1$ have evolved, while those with $qL(t) \ll 1$ are mostly frozen, retaining memory of the initial condition. As time evolves, $L(t)$ grows, and low-q modes have then in turn evolved. The two cases in Figs.\ref{fig:PinVsSinPin}(a) and (b) look similar (within the expected fluctuations for one sample). However, in the case with pinning $S(q,t)$ evolves more slowly than without pinning. This evolution becomes even slower as time/temperature increases/decreases during the temperature ramp. This is due to
vortices being trapped in local minima and the dynamics becoming thermally-activated. Moreover, at large times the structure has already relaxed much of the initial strain. Thus, the energy barriers separating metastable states have also grown with time, making the dynamics even slower. Although the system (with or without pinning)  strictly never stops evolving, the characteristic time-scale for equilibration grows significantly at low $T$ in the pinned case. We argue that this is also true in the experiment. Hence, in practice it is usual to refer to a ``freezing temperature'' to point out that the structure observed retains memory of a larger temperature/time because it can not be equilibrated at the much lower final temperature.

In order to prove, as argued in the main text, that in a diluted vortex system the main effect of pinning is to slow down the dynamics, in Fig.\ref{fig:PinVsSinPin}(b) we show that we can approximately match structure factors obtained with and without pinning. For instance, in the plot we show that $S_{\rm clean}(q,t=3) \approx S_{\rm pin}(q,t=4)$, $S_{\rm clean}(q,t=23) \approx S_{pin}(q,t=32)$, and $S_{\rm clean}(q,t=250) \approx S_{\rm pin}(q,t=512)$, where we have used $S_{\rm clean}(q,t)$ and $S_{\rm pin}(q,t)$ to denote the $S(q,t)$ obtained in simulations without and with pinning, respectively.

A hand-waving way to interpret why very dense weak pinning appears to be energetically the same as no pinning, is that pinned vortices can always reduce their  vortex-vortex interaction energy by jumping to one of the many available very close pinning centers, without appreciably changing its original pinning energy. This is valid as long as the vortex density is low enough, so pinning is not collective and individual jumps can not trigger other jumps (cf. Bragg Glass phase). Since jumps are thermal activated over finite pinning barriers, and such process is much slower than purely viscous diffusive motion at low temperatures, pinning only slows down the dynamics which minimizes the interaction free energy.
This picture is similar to the ``discrete'' superconductor idea proposed in the context low-field vortex systems with columnar defects \cite{vanderBeek2001,Colson2004}.

Therefore, the structure factor in the case with pinning can be approximately ``mapped'' to the one of the clean case at an earlier cooling time in the same protocol. This mapping allows to save computing time and to determine the temperature/time at which the clean system best mimics the experimental results.

\bibliography{biblio}

\end{document}